\documentclass[11pt,hidelinks,dvipsnames]{article}
\usepackage{authblk}
\usepackage[pdftex]{graphicx}
\title{Integrating protein sequence embeddings with structure via graph-based deep learning for single-residue property prediction
}

\author[1,*]{Kevin Michalewicz}
\author[1,*,$\dagger$]{Mauricio Barahona}
\author[1,*,$\dagger$]{Barbara Bravi}
\affil[1]{Department of Mathematics, Imperial College  London, London SW7 2AZ, United Kingdom}
\affil[*]{For correspondence: \texttt{\{k.michalewicz22, m.barahona, b.bravi\}@imperial.ac.uk}}
\affil[$\dagger$]{Equal contribution}

\date{}  
\usepackage[a4paper,headheight=16pt,scale={0.8,0.8},hoffset=0.5cm]{geometry}


\usepackage{float}
\usepackage[english]{babel}
\usepackage[utf8]{inputenc}
\usepackage{amsmath, bm}
\usepackage{multirow}
\usepackage{epigraph}
\usepackage[utf8]{inputenc}
\usepackage{makecell}

\renewcommand{\arraystretch}{1.4}
\usepackage{makecell}            
\usepackage{amssymb}
\usepackage{fancyhdr}
\usepackage{algorithm}
\usepackage{algpseudocode}
\usepackage{booktabs,siunitx}
\usepackage{array}
\usepackage[normalem]{ulem}
\usepackage{longtable}
\usepackage{csvsimple}
\usepackage{booktabs}
\usepackage{bbm}
\usepackage[bf]{caption}
\newcommand{\imgdir}{includes}
\graphicspath{{\imgdir/}}
\usepackage[T1]{fontenc}
\usepackage{titlesec, blindtext, color}
\usepackage{hyperref}
\definecolor{gray75}{gray}{0.75}
\newcommand{\hsp}{\hspace{20pt}}
\titleformat{\chapter}[hang]{\Huge\bfseries}{\thechapter\hsp\textcolor{gray75}{|}\hsp}{0pt}{\Huge\bfseries}
\usepackage{fontawesome}
\usepackage{subcaption}
\usepackage{pifont}
\usepackage{soul}
\usepackage[parfill]{parskip}
\usepackage{mathtools}
\newcommand{\angstrom}{\mbox{\normalfont\AA}}

\DeclareFontFamily{OT1}{cmbr}{\hyphenchar\font45 }
\DeclareFontShape{OT1}{cmbr}{m}{n}{%
  <-9>cmbr8
  <9-10>cmbr9
  <10-17>cmbr10
  <17->cmbr17
}{}
\DeclareFontShape{OT1}{cmbr}{m}{sl}{%
  <-9>cmbrsl8
  <9-10>cmbrsl9
  <10-17>cmbrsl10
  <17->cmbrsl17
}{}
\DeclareFontShape{OT1}{cmbr}{m}{it}{%
  <->ssub*cmbr/m/sl
}{}
\DeclareFontShape{OT1}{cmbr}{b}{n}{%
  <->ssub*cmbr/bx/n
}{}
\DeclareFontShape{OT1}{cmbr}{bx}{n}{%
  <->cmbrbx10
}{}

\newcommand*{\figref}[2][]{%
  \hyperref[{fig:#2}]{%
    Figure~\ref*{fig:#2}%
    \ifx\\#1\\%
    \else
      #1%
    \fi
  }%
}

\newcommand*{\tabref}[2][]{%
  \hyperref[{table:#2}]{%
    Table~\ref*{table:#2}%
    \ifx\\#1\\%
    \else
      #1%
    \fi
  }%
}

\newcommand*{\mtabrefs}[2][]{%
  \hyperref[{table:#2}]{%
    Tables~\ref*{table:#2}%
    \ifx\\#1\\%
    \else
      #1%
    \fi
  }%
}

\newcommand*{\mfigrefs}[2][]{%
  \hyperref[{fig:#2}]{%
    Figures~\ref*{fig:#2}%
    \ifx\\#1\\%
    \else
      #1%
    \fi
  }%
}

\newbox\keywbox
\setbox\keywbox=\hbox{\bfseries Keywords:}%

\newcommand\keywords{%
\noindent\rule{\wd\keywbox}{0.25pt}\\\textbf{Keywords:}\ }
\usepackage[numbers,super]{natbib}

\usepackage{xcolor}

\usepackage{cmbright}

\begin{document}

\maketitle

\begin{abstract}
\noindent Understanding the intertwined contributions of amino acid sequence and spatial structure is essential to explain protein behaviour. 
Here, we introduce INFUSSE (Integrated Network Framework Unifying Structure and Sequence Embeddings), a deep learning framework  for the prediction of single-residue properties that combines fine-tuning of sequence embeddings derived from a Large Language Model with the inclusion of graph-based representations of protein structures via a diffusive Graph Convolutional Network. 
%
To illustrate the benefits of jointly leveraging sequence and structure,
we apply INFUSSE to the prediction of B-factors in antibodies, a residue property that reflects the local flexibility shaped by biochemical and structural constraints in these highly variable and dynamic proteins.
Using a dataset of 1510 antibody and antibody-antigen complexes from the database SAbDab, we show that
INFUSSE improves performance over current machine learning (ML) methods based on sequence or structure alone, and allows for the systematic disentanglement of sequence and structure contributions to the performance. Our results show that adding structural information via geometric graphs enhances predictions especially for intrinsically disordered regions, protein-protein interaction sites, and highly variable amino acid positions---all key structural features for antibody function which are not well captured by purely sequence-based ML descriptions.

\keywords Antibody, Deep Learning, Graph-based Learning, Interpretability, Large Language Model, Protein Structure

\end{abstract}

\section{Introduction}
Protein function is the result of the complex interplay of amino acid sequence and how the polymer chain folds in three-dimensional space, leading to a characteristic, highly attuned physicochemical spatial structure.  
The development of sequencing technologies and the proliferation of large-scale repositories have resulted in a wealth of protein sequence data~\cite{UniProt}. There is also increasing accessibility to three-dimensional protein structures, available through the Protein Data Bank (PDB)~\cite{pdb}. Bringing together both sources of information remains a critical area of study~\cite{Varadi2022}, specifically to determine how and in which contexts structural data can significantly improve the prediction of protein properties, both global or at the single-residue level.

Recent advances in Large Language Models (LLMs) have revolutionised protein modelling and design by harnessing the ability of transformer architectures to capture complex patterns and long-range dependencies within protein sequences. Previously, Recurrent Neural Networks (RNNs) had struggled with exploding and vanishing gradients---hence not useful for long sequences~\cite{Schmidt2019, Bengio1994}. Similarly, Long Short-Term Memory networks (LSTMs)~\cite{Hochreiter1997} were still limited to a finite-context window---hence only helpful to capture short- and medium-range sequence correlations. In contrast, transformers utilise self-attention to process the entire sequence at once and are able to capture long-range dependencies efficiently~\cite{Vaswani2017}. A leading example of the transformer architecture for protein analysis is ProtBERT~\cite{Elnaggar2022}, which was trained on more than 200 million protein sequences, demonstrating high performance for sequence-based predictions~\cite{Littmann2021, Marquet2022}. An important feature of ProtBERT is that it generates sequence 
embeddings, \textit{i.e.}, enriched sequence representations with enhanced biological information relative to simple one-hot encoding of amino acid types. Such embeddings have been fruitfully exploited to facilitate and increase the performance of downstream prediction tasks~\cite{gligorijevic_structure-based_2021, Bepler2019, Ibtehaz2023}.  
 
Although LLMs have demonstrated potential to predict protein properties from sequence data alone, the addition of structural information often unlocks deeper insights and enhances performance~\cite{Shu2024}. Structural information can be effectively represented using graphs, where nodes correspond to atoms or residues and edges encode interactions or geometric relationships. Atomistic graphs have been employed to identify rigid and flexible regions within proteins~\cite{Jacobs2001, delvenne2010, delmotte2011}, unravel allosteric mechanisms~\cite{amor_prediction_2016, Dubanevics2022,Wu2022}, and deepen our understanding of protein-protein interactions~\cite{Jha2022}. Coarse-grained graphs at the residue level have also proven effective in various applications~\cite{bahar_global_2010}, including protein folding~\cite{Liwo2005,Kmiecik2012,Maisuradze2010}, aggregation studies~\cite{Nasica-Labouze2015}, protein-protein docking~\cite{Fleishman2010}, allosteric functional modulation~\cite{DelSol2009} and protein flexibility prediction~\cite{Frezza2015}.

Recently, deep learning architectures such as Graph Convolutional Networks (GCNs)~\cite{Kipf2017} have been introduced to use graphs to exploit relational information in data for machine learning (ML) tasks. Taking a graph and a set of feature-based descriptions (embeddings) of the nodes as input, GCNs update the embeddings via graph convolutions to enhance node prediction by accounting for the relationships modelled by the graph. 
%
In the context of ML for protein modelling, GCNs have been leveraged to classify proteins into functional families~\cite{gligorijevic_structure-based_2021, Jiao2023, LiuPolat2024}, predict protein-protein interactions~\cite{Gao2023}, and estimate protein-ligand binding affinity~\cite{Zhang2023}. However, with the exception of Ref.~\cite{Canner2023}, which identifies carbohydrate-binding sites, a residue-level property, using structural data alone, these approaches are designed to predict \emph{global} protein properties.
%
%
Our aim here is   
to integrate systematically sequence-based with graph-aware \emph{residue-level} descriptions 
and to assess quantitatively the contribution of sequence and graph-based structural information for residue-level prediction tasks.
%

To this end, we introduce INFUSSE, a deep learning framework that, starting from task-agnostic LLM sequence embeddings, generates enriched, task-driven sequence representations, and integrates them with graph representations of protein structures for the learning of residue-specific properties within proteins. The graph integration step relies on a \textit{diffusive} GCN (diff-GCN)~\cite{Peach2020}, a recent version of GCNs that uses a diffusive process to propagate information across the molecular graph at learnable scales. INFUSSE is a general framework for node prediction, and thus applicable to any single-residue property.  

We illustrate our framework through the prediction of local residue flexibility (B-factors) of antibody-antigen complexes. B-factors (also known as Debye-Waller or temperature factors) measure how much atoms fluctuate around their average position at equilibrium, \textit{i.e.}, a higher B-factor corresponds to greater thermal motion at a given temperature at a particular site~\cite{Bramer2018}. B-factors provide insights into protein dynamics and the link between structure and function~\cite{Sun2019}, as intrinsic flexibility correlates with conformational changes~\cite{Ma2005, Seoane2021} and has been proposed to be informative about the propensity to form protein binding interfaces~\cite{Seoane2021}.

Several methods have been proposed for the prediction of B-factors for general proteins. Ref.~\cite{Xia2014} used Gaussian Network Models (GNMs)~\cite{Rader2005} and persistent homology concepts. Ref.~\cite{Bramer2018} applied various ML techniques to features derived from PDB data and achieved a maximum Pearson correlation coefficient (PCC) $R$ between ground truth and predicted B-factors of $0.66$. 
Recently, Ref.~\cite{Pandey2023} applied a bidirectional LSTM to input protein data comprising atomic spatial coordinates, primary sequence, secondary structure, and chain-break information to obtain state-of-the-art (SOTA) performance with $R=0.8$ on general proteins. However, as part of our comparison, we have found that the application of this LSTM architecture to antibodies, which are proteins with flexible, unstructured and highly variable regions, achieves noticeably lower performance ($R=0.48$). In this work, we develop the INFUSSE architecture and benchmark its performance in connection to B-factor prediction in antibodies by comparing it to such other methods in the literature and to variants of our own architecture that include only sequence or structure inputs. Specifically, we perform a systematic evaluation of the contribution of sequence and structure to the quality of the prediction. This comparison allows us to pinpoint scenarios where using the sequence alone faces limited predictability and the inclusion of structural graph information leads to enhanced predictive power, \textit{e.g.}, in regions characterised by high sequence variability. 

Our work focuses on antibody and antibody-antigen complexes, since antibodies are prototypes of proteins containing highly variable and unstructured regions (primarily the Complementarity Determining Regions, CDRs), for which the prediction of B-factors remains challenging. Both the diff-GCN and LLM architectures employed by INFUSSE are alignment-free, which is particularly convenient when working with protein sequences of considerable length variability such as antibodies, and immune receptors more generally. Furthermore, the choice of antigen-bound complexes allows us to analyse the prediction of B-factors at the binding interfaces, a main target for the study of immune escape mutants and for the design of antigen-specific antibodies.

\section{Results}
\subsection*{INFUSSE is designed to leverage jointly sequence and structure information for residue-specific predictions}
\label{sec:infusse}
\begin{figure}[htb!]
\centering
\includegraphics[width=\linewidth]{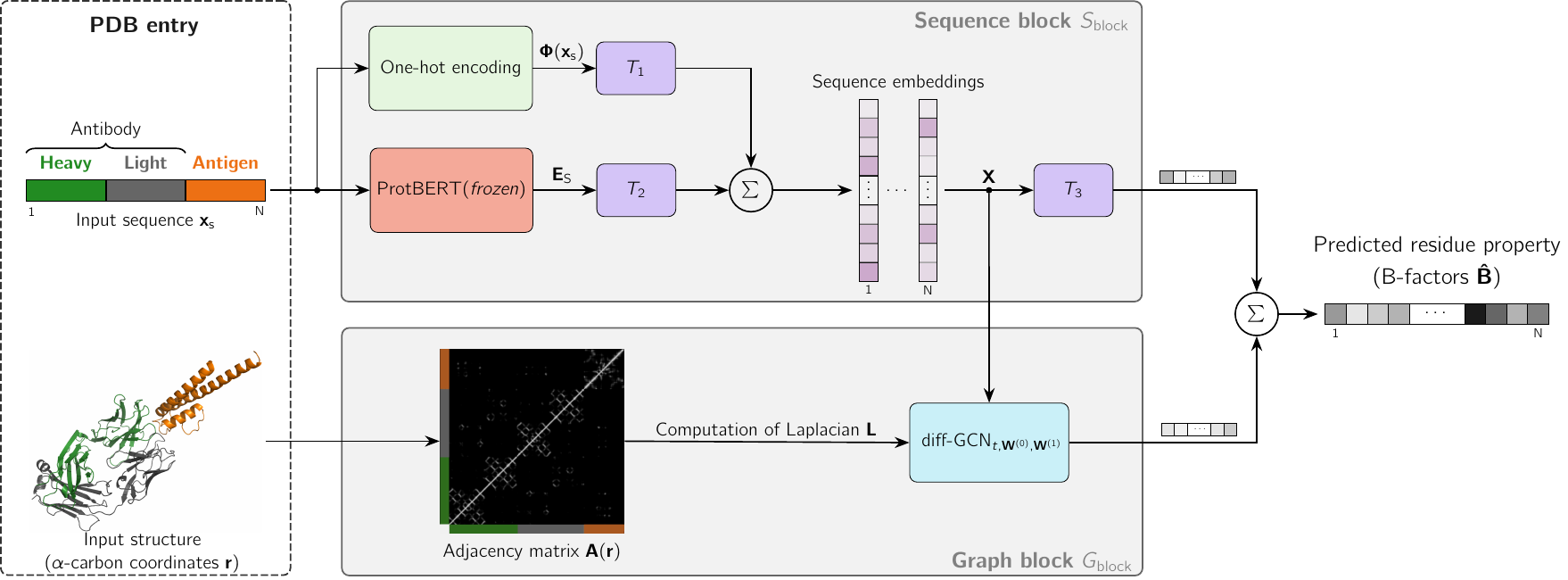}
\caption{\textbf{INFUSSE architecture}. A frozen LLM (ProtBERT) and a diffusive Graph Convolutional Network (diff-GCN) are combined to predict B-factors for antibody-antigen complexes. Input sequences are encoded by a frozen ProtBERT model and passed through  $T_2$, then summed with their one-hot encoded version transformed by $T_1$, producing embeddings that are further transformed by $T_3$. $T_1$, $T_2$ and $T_3$ are learnable non-linear transformations.
The resulting representations are summed with the output of the diff-GCN, with learnable parameters $t$, $\mathbf{W}^{(0)}$ and $\mathbf{W}^{(1)}$, which takes the enriched sequence representation $\mathbf{X}$ from $S_{\mathrm{block}}$ as input node features and leverages the Laplacian of a geometric graph constructed from the $\alpha$-carbon 3D coordinates of the antibody-antigen complex.}
\label{fig:gcn-bf}
\end{figure}

We start by introducing INFUSSE (Integrated Network Framework Unifying Structure and Sequence Embeddings), a deep learning architecture for protein modelling designed to perform residue-specific property prediction by integrating protein sequence embeddings and protein structural graphs through a sequence block $S_{\mathrm{block}}$ and a graph block $G_{\mathrm{block}}$ (Figure~\ref{fig:gcn-bf}, Methods, Section \ref{sec:full_infusse}). As our use case, we apply this architecture for the prediction of single-residue B-factors, which are central to the quantitative characterisation of the conformational flexibility of a residue, as determined by the molecular packing and biophysical interactions within its structural neighbourhood. Therefore, this task showcases a typical scenario where integrating structural information via graphs can be essential. Following Ref.~\cite{Pandey2023}, we predict \textit{standardised} B-factors (zero mean, unit variance per PDB entry), as they reflect the dynamics of different protein regions in a more comparable way, without hinging upon the resolution of the structure as a predictor, in contrast to other works~\cite{Bramer2018}.

The sequence block of INFUSSE, $S_{\mathrm{block}}$, takes as input the length $N$ sequence
$\mathbf{x}_{\mathrm{s}}$ of a protein or a protein complex (here an antibody or antibody-antigen complex) and produces an enriched, task-driven representation of it composed of multiple features, $\mathbf{X}$, by: computing, through a pre-trained, well established LLM (ProtBERT~\cite{Elnaggar2022}), an LLM embedding of it  $\mathbf{E}_{\mathrm{s}}:=\mathrm{ProtBERT(\mathbf{x}_{\mathrm{s}})}$; passing $\mathbf{\Phi}(\mathbf{x}_{\mathrm{s}})$, a one-hot encoded version of $\mathbf{x}_{\mathrm{s}}$, and $\mathbf{E}_{\mathrm{s}}$ through two transformations $T_1$ and $T_2$ parametrised by non-linear layers and learnt for residue-specific property prediction; summing the output of $T_1$ and $T_2$. 

The resulting sequence representation $\mathbf{X} \left(\mathbf{x}_{\mathrm{s}}, \mathbf{E}_{\mathrm{s}} \right)$ leverages the biologically relevant, sequence context-aware patterns captured by the LLM embeddings, but is further fine-tuned on the specific protein data and prediction task of interest via $T_1$ and $T_2$. $\mathbf{X}$ is passed as input to an additional learnable transformation $T_3$, to obtain the $S_{\mathrm{block}}$ prediction, \textit{i.e.}, a purely sequence-based prediction of the target residue-specific property.

$G_{\mathrm{block}}$ starts from the spatial coordinates $\mathbf{r}$ of the $N$ $\alpha$-carbons of the input protein (or protein complex) and builds a graph representation of its structure specified by the graph adjacency matrix $\mathbf{A}(\mathbf{r})$, which models interactions between residues at different positions as graph edges. Different schemes exist to generate graph descriptions from protein structural data: whereas some methods use both structural data and detailed energetics of physicochemical potentials~\cite{Song2021, Wu2022_2}, simpler \emph{geometric} constructions rely only on Euclidean distances between residues~\cite{Liu2020}. Given that our aim is to predict residue conformational flexibility, which is related to atomic packing, we focus here on \emph{geometric} graph constructions. Specifically, we considered the geometric construction of \textit{weighted Gaussian graphs}, whereby the interaction strength between residues decays as a Gaussian function of their distance, hence its adjacency matrix $\mathbf{A}(\mathbf{r})$ (depicted in Figure \ref{fig:gcn-bf}) summarises residue spatial vicinity. As a term of comparison we considered also the geometric construction of a Gaussian Network Model (GNM), which sets an edge in the graph only between residues that are spatially closer than a cut-off distance (see Methods, Section \ref{sec:protein_graphs} for definitions). 

Given $\mathbf{A}(\mathbf{r})$, $G_{\mathrm{block}}$ computes the so-called \emph{graph Laplacian} (Methods, Section \ref{sec:protein_graphs}), a matrix that encodes the diffusive dynamics on a given graph, and can be used here to `diffuse' structurally relevant information within a neighbourhood of closely packed residues to predict effectively single-residue B-factors. This is done through the diff-GCN (Methods, Section \ref{sec:diffGCN}), where the feature-based representation of single residues sitting at the graph nodes is initialised to $\mathbf{X}$. The learnable parameters of the diff-GCN are two weight matrices $\mathbf{W}^{(l)}$ with $l=0,1$ between node embedding features and a parameter $t$ these embeddings which sets the scale of the information-diffusive step between nodes (see Eq.~\eqref{eq:diffGCN} in Methods). 

The outputs of the sequence and graph blocks are summed to yield the final output by INFUSSE (Figure~\ref{fig:gcn-bf}, Methods, Section \label{sec:full_infusse}), \textit{i.e.}, the set of predicted standardised B-factors $\hat{\mathbf{B}}= \{ \hat{B}_j \}$ for each residue $j$. Such a prediction, given as inputs the one-hot encoded protein sequence $\mathbf{x}_{\mathrm{s}}$, its representation $\mathbf{E}_{\mathrm{s}}$ obtained through LLM embeddings and the residues' spatial coordinates $\mathbf{r}$, is computed as:
\begin{equation}
\label{eq:full_infuse}
\hat{\mathbf{B}} = \text{INFUSSE}(\mathbf{x}_{\mathrm{s}}, \mathbf{r}):= S_{\mathrm{block}} \left(\mathbf{x}_{\mathrm{s}}, \mathbf{E}_{\mathrm{s}} \right) + G_{\mathrm{block}}(\mathbf{r}, \mathbf{X} (\mathbf{x}_{\mathrm{s}}, \mathbf{E}_{\mathrm{s}}))
\end{equation}

In conclusion, as indicated by Eq.~\eqref{eq:full_infuse}, INFUSSE integrates the two main information channels of protein data (sequence and structure) by optimising both the purely sequence-based residue embeddings (from $S_{\mathrm{block}}$) and the graph-aware ones (from $G_{\mathrm{block}}$) toward a residue-specific prediction task, hence leveraging them synergistically and allowing us to assess systematically their role in prediction.

\subsection*{INFUSSE outperforms generic protein SOTA models for B-factor prediction in antibody-antigen complexes}
To train INFUSSE in this work, we retrieved a total of $1510$ high-quality structures from the PDB via query of the Structural Antibody Database (SAbDab)~\cite{SAbDab2014,SAbDab2022} (Methods, Section \ref{sec:data}), comprising $1143$ antibody-antigen complexes and $367$ unbound antibodies with B-factor annotation. We produced 10 splits of these $1510$ PDB entries into training and test sets, containing $1435$ (95\%) and $75$ (5\%), respectively, for each split.

From a graph learning perspective, the prediction of a residue-level property is a node regression task. As such, training consists of the minimisation of a mean-squared error of B-factor prediction and is conducted in a \textit{two-step} manner (Methods, Section \ref{sec:training}). In a first step, we train the sequence block $S_{\mathrm{block}}$ by learning the parameters of $T_1$, $T_2$ and $T_3$ to predict B-factors from sequence alone, \textit{i.e.}, from the sequence information $\mathbf{x}_{\mathrm{s}}$ and the LLM (ProtBERT) embedding $\mathbf{E}_{\mathrm{s}}$. In a second step, we incorporate the graph block $G_{\mathrm{block}}$ and we optimise the parameters of the diff-GCN ($t$, $\mathbf{W}^{(l)}$) \textit{jointly} to the ones of $T_1, T_2, T_3$ (initialised to the result of the $S_{\mathrm{block}}$ training from the first step). ProtBERT is kept frozen, and only used to compute informative sequence embeddings.

After training, the model was evaluated on the test set by computing the Pearson correlation coefficient $R$ between the ground truth and predicted standardised B-factors, repeating training and testing for 10 independent training/test data splits (Figure 2B and Methods, Section \ref{sec:methods_evaluation}). 

\begin{table}[h!]
\centering
\footnotesize
\caption{\textbf{Performance of different methods for the prediction of standardised B-factors of residues of antibody–antigen complexes.}
The measure of performance reported is the Pearson correlation coefficient $R$, averaged over 10 training/test splits, with standard deviation across splits.}
\resizebox{\textwidth}{!}{
\begin{tabular}{|l|c|c|c|c|}
\hline
\textbf{Method} &
\textbf{\makecell{Sequence \\ representation}} &
\textbf{\makecell{Structure \\ representation}} &
\textbf{\makecell{Learnt with ML}} &
$R$ \\ 
\hline

INFUSSE ($S_{\mathrm{block}}$ + $G_{\mathrm{block}}$) &
\makecell{One-hot encoding \\ \& LLM embeddings} &
Weighted Gaussian graph &
\makecell{$T_1, T_2, T_3, \mathbf{W}^{(l)}, t$} &
$0.71 \pm 0.01$ \\
\hline

$S_{\mathrm{block}}$ alone (no structure) &
\makecell{One-hot encoding \\ \& LLM embeddings} &
--- &
\makecell{$T_1, T_2, T_3$} &
$0.64 \pm 0.02$ \\
\hline

\makecell[l]{$S_{\mathrm{block}}$ (no LLM) + $G_{\mathrm{block}}$} &
One-hot encoding &
Weighted Gaussian graph &
\makecell{$T_1, T_3, \mathbf{W}^{(l)}, t$} &
$0.55 \pm 0.04$ \\
\hline\hline
\makecell[l]{LSTM~\cite{Pandey2023} (SOTA general proteins)} &
One-hot encoding &
\makecell{Raw coordinates $\mathbf{r}$,  secondary \\ structure, and chain breaks} &
LSTM weights &
$0.48 \pm 0.06$ \\
\hline

\makecell[l]{Laplacian pseudoinverse (no learning)} &
--- &
Weighted Gaussian graph &
--- &
$0.01 \pm 0.04$ \\
\hline
\end{tabular}}
\label{table:gcn-bf-perf}
\end{table}

INFUSSE achieved $R=0.71 \pm 0.01$, averaged over the 10 training/test splits, for the prediction of standardised B-factors (see~\figref[B]{2} and Table~\ref{table:gcn-bf-perf}). This performance was achieved with a weighted Gaussian graph~(Eq.~\eqref{eq:Gaussiangraph} in Methods with $\eta=8$), while INFUSSE with GNM graphs (Eq.~\eqref{eq:GNM} in Methods with $\epsilon=10\angstrom$) had only slightly lower performance ($R=0.70$), and similarly when a GCN without diffusion was used (see Table S1 for a full comparison of these versions). These consistent results support the robustness of the INFUSSE architecture.

Table~\ref{table:gcn-bf-perf} also shows the results of two other models that follow a \textit{partial} INFUSSE architecture, \textit{i.e.}, they include only some of the ingredients of the full INFUSSE model. The `$S_{\mathrm{block}}$ alone' model is our step 1 model trained only on sequence and ProtBERT embeddings, \textit{i.e.}, $S_{\mathrm{block}} (\mathbf{x}_{\mathrm{s}}, 
 \mathbf{E}_{\mathrm{s}} )$, and it achieved an average $R=0.64$. The `$S_{\mathrm{block}}$ (no LLM) + $G_{\mathrm{block}}$' model corresponds to $S_{\mathrm{block}} \left(\mathbf{x}_{\mathrm{s}}, \mathbf{0} \right) +
G_{\mathrm{block}}(\mathbf{r}, \mathbf{X} (\mathbf{x}_{\mathrm{s}}, \mathbf{0})
)$, \textit{i.e.}, ProtBERT embeddings are not used but otherwise sequence and geometric graphs are employed, and it achieved an average $R=0.55$.
All together, these results underscore the importance of including both sequence embeddings and structural graphs within INFUSSE for the prediction of B-factors. 

In Table~\ref{table:gcn-bf-perf}, we also compared INFUSSE to two models from the literature that perform well for generic proteins.
The SOTA LSTM model for B-factor prediction in general proteins~\cite{Pandey2023} produced an average $R=0.48$ when applied to our antibody-antigen dataset, in contrast to the $R$ between $0.6-0.8$ that it achieved on generic protein benchmarks~\cite{Bramer2018, Pandey2023}. This reflects the difficulty of B-factor prediction for antibodies due to their high sequence variability and the presence of unstructured regions.  
Similarly, a baseline graph model that relies on the pseudoinverse of the graph Laplacian with no learning~\cite{bahar_global_2010} produced $R= 0.01$ for both weighted Gaussian graphs and GNMs on our antibody dataset, in contrast to $R$ between $0.65$ and $0.8$ on generic proteins~\cite{Rader2005}. Again, this underscores the challenges of B-factor prediction in antibody-antigen complexes, and the need to combine enriched sequence representations and structural graph information, as in our INFUSSE model. 

\subsection*{INFUSSE disentangles sequence and structure contributions to B-factor prediction}
A focus of our work is to identify the protein regions, structural motifs and positions where the inclusion of structural information via graphs influences predictive performance. To characterise the scenarios in which the addition of a structural graph significantly affects B-factor prediction, we considered the errors in predicting B-factors using only sequence information, as outputted by $S_{\mathrm{block}}$:
\begin{equation}
\label{eq:epsilon_Sblock}
\varepsilon^{(q)}_{S_{\mathrm{block}}, j} := \left(\hat{B}^{(q)}_{S_{\mathrm{block}},j} - B^{(q)}_j \right)^2
\end{equation}
for a given input sample $q$ and residue with position $j$, and we compared it to the error of the full INFUSSE model:
\begin{equation}
\label{eq:epsilon_infusse}
\varepsilon^{(q)}_{\text{INFUSSE}, j} := \left(\hat{B}^{(q)}_{j} - B^{(q)}_j \right)^2 
\end{equation}
where both sequence embeddings and graph are present (Eq.~\eqref{eq:full_infuse}). 

\figref[B]{2} shows the comparison between these two prediction errors across all samples and their residues. In most cases, the error of the INFUSSE model is smaller than the error of $S_{\mathrm{block}}$, highlighting the benefits of including structural information in B-factor prediction. To better quantify 
this effect across protein regions and motifs, we defined the differential prediction errors: 
\begin{equation}
\label{eq:delta_graph_eq_main}
\Delta^{(q)}_{\text{graph}, j} := \varepsilon^{(q)}_{S_{\mathrm{block}},j} -\varepsilon^{(q)}_{\text{INFUSSE},j} 
\end{equation}
for residues at position $j$ in the input sample $q$. (We will use primarily the more compact notation $\Delta_{\mathrm{graph}}$ to denote sets of $\Delta^{(q)}_{\text{graph}, j}$ values across different positions $j$ and samples $q$, see Methods, Section \ref{sec:delta_graph} for more details). 
$\Delta_{\mathrm{graph}}$ measures the magnitude of the graph-induced change in predictive power of INFUSSE compared to a purely sequence-based prediction leveraging LLM embeddings, and captures a graph-induced improvement when  $\varepsilon^{(q)}_{\text{INFUSSE}, j} < \varepsilon^{(q)}_{S_{\mathrm{block}}, j}$.

We next assessed whether the graph-induced differential performance, measured by $\Delta_{\text{graph}}$ at the residue-level, follows systematic patterns across several biological scenarios; namely, in relation to  positions with greater sequence variability, unstructured regions, and binding interfaces.
To this end, we evaluated $\Delta_{\text{graph}}$ on the test set, stratifying its pattern by such scenarios, and we quantified statistical differences using three complementary statistical tests: (i) a one-sample test in which we try to reject that the group mean of $\Delta_{\text{graph}}$ is equal to zero, to indicate if the graph-induced differential performance corresponds to an \emph{improvement}; (ii) a two-sample comparison of means testing whether the difference of means $\Delta\mu$ between two groups (\textit{e.g.}, high- \textit{vs} low-entropy, CDR \textit{vs} FR, etc.) differs from zero, to flag up the scenarios for which the graph-induced differential performance is more enhanced compared to other ones; and (iii) a two-sample comparison of interquartile ranges (IQRs) testing whether their difference $\Delta\mathrm{IQR}$ between two groups differs from zero, to flag up the scenarios for which the graph-induced differential performance is more heterogeneous (Methods, Section \ref{sec:statistical_tests}).

\begin{figure}[h!]
\centering  \includegraphics[clip, trim=0cm 0cm 6cm 0cm, width=\linewidth]{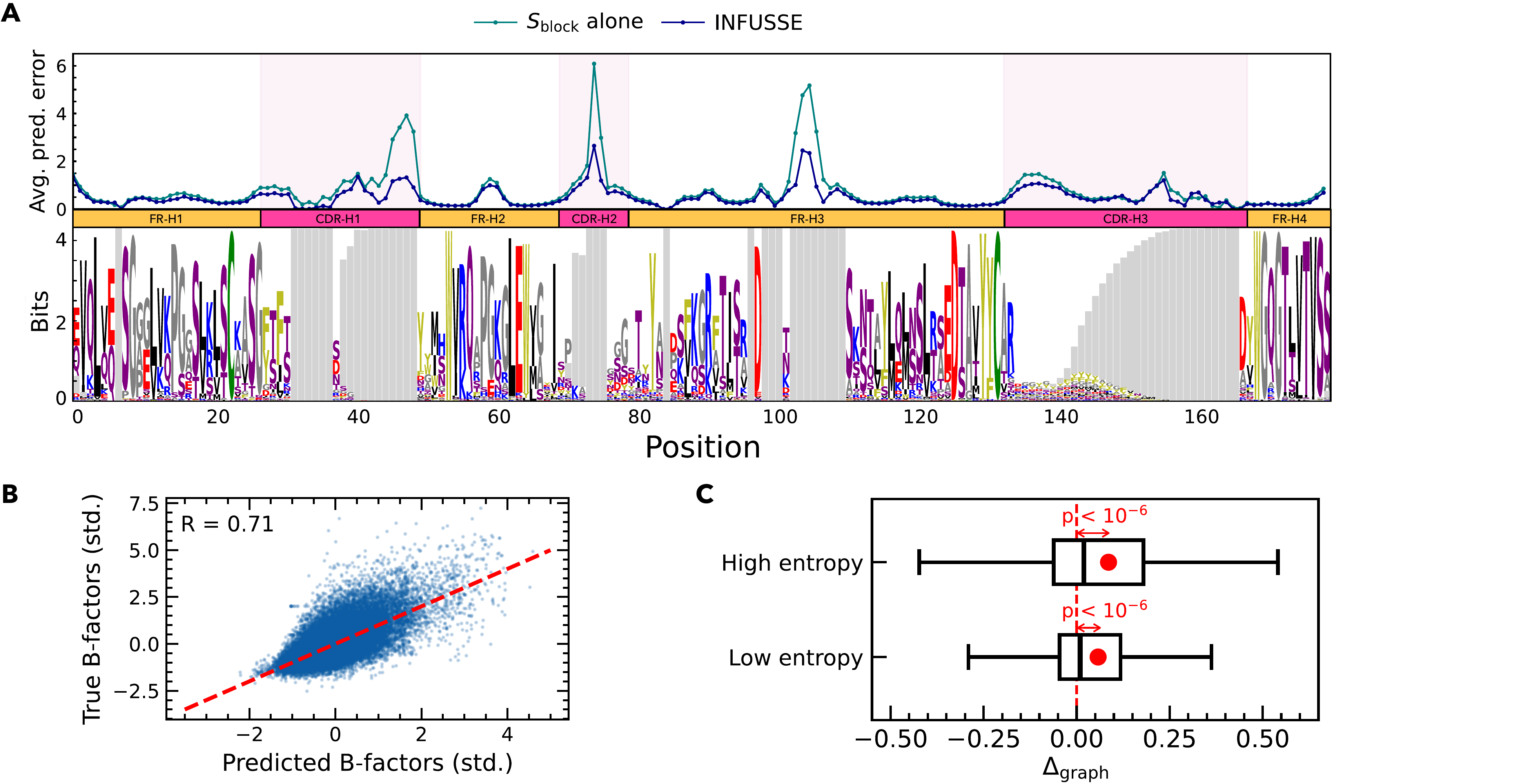}
  \caption{\textbf{INFUSSE model performance.} (\textbf{A}) Top - Prediction errors of INFUSSE, $\varepsilon_{\text{INFUSSE}, n}^{(q)}$, and $S_{\mathrm{block}}$ alone, $\varepsilon_{S_\mathrm{block}, n}^{(q)}$, for each position $n$ of the heavy chain antibody variable region (aligned with ANARCI) averaged over all the samples $q$ (see Figure~\ref{fig:S1} for the corresponding light chain results). Pink shaded areas indicate CDRs. Bottom - Sequence logos of aligned heavy chain variable region for the entire dataset, where the grey boxes denote missing amino acids. (\textbf{B}) INFUSSE predictions of standardised B-factors on the test set for one of the training/test splits with test-set $R =0.71$. (\textbf{C}) Boxplots of the set $\Delta_{\mathrm{graph}, j}^{(q)}$ of graph-induced differential performance, where $j$ extends to all antibody variable region sequence positions. The positions are then divided into high and low entropy groups based on their amino acid diversity (see A, bottom, and Methods, Section \ref{sec:scenarios}). Red dots indicate mean values and black vertical lines denote medians. 
  }
\label{fig:2}
\end{figure}

\subsubsection*{Graph-induced improvement correlates with high sequence diversity}

First, we considered the relationship between graph-induced differential performance and sequence composition diversity at each position. To do so, we aligned the entire dataset of antibodies using ANARCI~\cite{ANARCI} (Methods, Section \ref{sec:scenarios}), we computed the position-specific prediction error of the INFUSSE model (Eq.~\eqref{eq:epsilon_infusse}) averaged over the whole antibody dataset and compared it to the average prediction error of the purely sequence-based prediction from $S_{\mathrm{block}}$ (Eq.~\eqref{eq:epsilon_Sblock}). We found a graph-induced improvement in performance (lower INFUSSE error) at most positions along the antibody sequence (see \figref[A]{2} and Figure~\ref{fig:S1}).

By considering the entropy at each position, indicated by the sequence logos in \figref[A]{2} (bottom), we found noticeably higher graph-induced improvement 
in positions with high residue diversity, including insertion positions where only a limited number of antibodies contain an amino acid (grey bars).
To assess this association, we analysed the relation of the differential prediction errors $\Delta_{\mathrm{graph}}$ over all positions to their entropy-based diversity score (Eq.~\eqref{eq:diversity_scores_set} in Methods). \figref[C]{2} shows statistically significant differences in $\Delta_{\text{graph}}$ when comparing positions with high and low entropies ($\Delta\mu=0.03$, p-value $3\times10^{-5}$) and IQRs ($\Delta\mathrm{IQR}=0.08$, p-value $<10^{-6}$), see Methods, Section \ref{sec:statistical_tests} and \tabref[A]{S2} for details. 

These results indicate that the inclusion of graph information is more beneficial for regions with higher residue diversity. Such high-entropy positions are not only associated with greater average improvement from the graph, but also with a broader interquartile range (IQR), suggesting more heterogeneous graph-induced prediction differences. This reflects the fact that such positions (and molecular environments) are naturally undersampled in the dataset, thus posing a specific challenge to sequence-based learning, as seen in the large average and IQR of the errors of the `$S_{\mathrm{block}}$ alone' model, $\varepsilon_{S_\mathrm{block}}$, presented in 
\figref[A]{S3}. Importantly, such high variability is intrinsic to the functionality of antibodies, especially within the CDRs.  
Consistent with these observations, Figure~\ref{fig:S1} shows that the average graph improvement is smaller for the light chain of the antibody, which is known to display lower variability~\cite{Jaffe2022}.

\subsubsection*{Graph-induced improvement correlates with CDR regions and less ordered secondary structures}

We next assessed the association of the graph-induced differential performance $\Delta_{\mathrm{graph}}$ with two aspects of antibody and protein structure. Firstly, we examined the relation to residues being part of the  Framework (FR) region compared to being part of the Complementarity-determining regions (CDRs) in the antibody variable region. CDRs are known for their structural plasticity and sequence variability~\cite{Akbar2021}, which enable the immune system to recognise a diverse range of antigens. Such variability is clearly noticeable in the sequence diversity of \figref[A]{2} (pink shaded areas), hence we expect graphs to induce higher improvement for B-factor prediction in such regions.
Our results in~\figref[A]{3} and \tabref[B]{S2} show significantly higher mean and IQR of $\Delta_{\mathrm{graph}}$ values for residues in CDRs compared to those in FRs ($\Delta\mu=0.08$, p-value $< 10^{-6}$; $\Delta\mathrm{IQR}=0.15$, p-value $< 10^{-6}$). The mean $\Delta_{\mathrm{graph}}$ is significantly positive for both FR ($\mu=0.05$, p-value   $< 10^{-6}$) and CDR ($\mu=0.09$, p-value $< 10^{-6}$) regions, indicating that employing graph information leads on average to an improvement especially for the entire antibody variable region. The larger variability (higher IQR) observed in the CDRs follows from the fact that regions with high sequence variability, dynamic behaviour and intrinsic disorder lead to variable and poor performance of the $S_\mathrm{block}$ alone 
(\figref[B]{S3}, Table~\ref{table:S3}).

\begin{figure}[h!]
    \centering
    \includegraphics[clip, trim=0cm 0.cm 12cm 0cm, width=\linewidth]{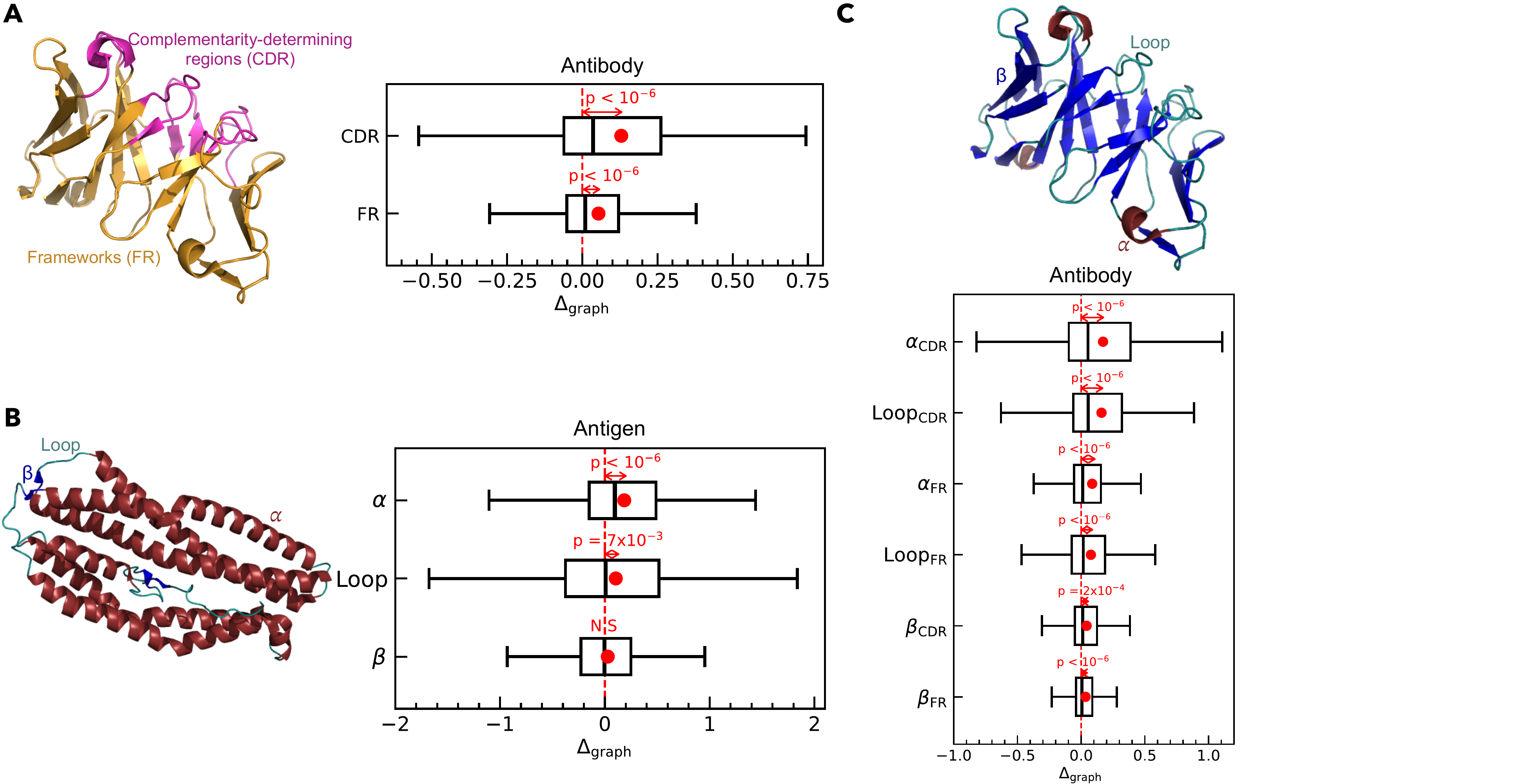}
\caption{\textbf{Graph-induced differential performance stratified by structural motifs.} Structures from PDB entry \texttt{6wm9} are included as illustrations of the statistical analyses in \textbf{A}, \textbf{B}, \textbf{C}. (\textbf{A}) Boxplot of graph-induced differential performance ($\Delta_{\mathrm{graph}}$) for antibody amino acids belonging to Framework (FR) versus Complementarity-determining regions (CDR) over the test set. (\textbf{B}) Boxplot of $\Delta_{\mathrm{graph}}$ stratified by secondary structure types ($\alpha$-helix, $\beta$-strand or loop) for the antigens in the test set. (\textbf{C}) Boxplot of $\Delta_{\mathrm{graph}}$ for residues in antibody variable regions from the test set, stratified by secondary structure type as well as membership to CDR or FR.}
\label{fig:3}
\end{figure}

Secondly, we analysed different secondary structure motifs ($\alpha$-helices, $\beta$-strands and loops) across antibody and antigen.
For the antigen residues, we found that the graph information brings an improvement for $\alpha$-helices and loops, but shows no significant change compared to sequence-based prediction alone for $\beta$-strands (see~\figref[B]{3} and \tabref[C]{S2}). We found a significantly higher $\Delta_{\mathrm{graph}}$ mean value for $\alpha$-helices compared to loops ($\Delta\mu=0.08$, p-value   $0.05$), while the latter's IQR is wider than that of helices ($\Delta\mathrm{IQR}=0.25$, p-value   $10^{-6}$), suggesting more heterogeneity of graph-induced effects on performance across positions. These findings aligns with our intuition. 
As shown in \figref[C]{S3} and Table~\ref{table:S3},
the intrinsic flexibility and conformational variability of \emph{unstructured} regions make the mapping between sequence, structure and B-factors less consistent in the training dataset, leading to more heterogeneous degrees of performance. Similarly, the narrow, near-zero distribution for $\beta$-strands indicates that the sequence block already captures most of their geometry and conformational flexibility, which are encoded to a large extent in their amino acid composition. $\beta$-strands are indeed particularly regular structural motifs whose stability is enabled by specific patterns of hydrophobic residues and the hydrogen bonds they form~\cite{Wouters1995}.

For the antibody residues (see~\figref[C]{3} and \tabref[D]{S2}), we found that $\alpha$-helices and loops within CDRs behave similarly, both displaying large positive mean values and broad IQRs, whereas the same motifs in FRs yield significantly lower means and narrower spreads. 
The wider IQR of $\alpha$-helices within CDRs, compared to loops, is again structural: $\alpha$-helices are typically much shorter in antibodies than in antigens (see~\figref[]{S2}), they are not present in the CDRs of all antibodies, and when they do occur, they tend to be more variable and exposed to the antigen in CDRs, as they play a critical role in binding~\cite{Lowe2011}. 
Graphs contribute little in either FR or CDR regions for $\beta$-strands, though $\beta$-strands in the CDR retain a slightly wider IQR than those in the FR. The enrichment of conserved hydrophobic residues in $\beta$-strands of FR regions supports the idea that the conformation and stability of $\beta$-strands are highly sequence-dependent~\cite{Abeln2014}, while also helps to explain why these conserved structural elements tend to have lower $\Delta_{\mathrm{graph}}$ values. The three regimes, \textit{i.e.}, $\alpha_\mathrm{CDR}$/$\mathrm{Loop}_\mathrm{CDR}$, $\alpha_\mathrm{FR}$/$\mathrm{Loop}_\mathrm{FR}$ and $\beta_\mathrm{CDR}$/$\beta_\mathrm{FR}$, underscore that flexible, surface-exposed motifs like CDRs benefit most from explicit structural context. In contrast, regular, conserved elements (for instance FR $\beta$-strands) are already well-accounted for by $S_{\mathrm{block}}$ alone. 
See~\mfigrefs[C,D]{S3} and Table~\ref{table:S3} for detailed errors of the INFUSSE and $S_{\mathrm{block}}$ models for the different structural motifs across antigen and antibody regions.


\subsubsection*{Graph information enhances prediction for paratope and epitope members}

Sequence-based approaches often struggle to accurately describe the properties of paratope and epitope members (\textit{i.e.}, the interfacial residues between antibody and antigen), due to the inability of such approaches to capture, from sequence alone, how chains fold, and therefore how they interact within the antibody-antigen complex~\cite{Hou2021}. This is confirmed in our B-factor prediction task. \figref[A]{4} illustrates in one structure $q$ of the test set how high values of $\Delta_{\mathrm{graph}}^{(q)}$ characterise primarily paratope and epitope members, indicating high improvement achieved by adding the graph in those regions.

\begin{figure}[h!]
    \centering
    \includegraphics[clip, trim=0cm 0cm 2.3cm 0cm, width=\linewidth]{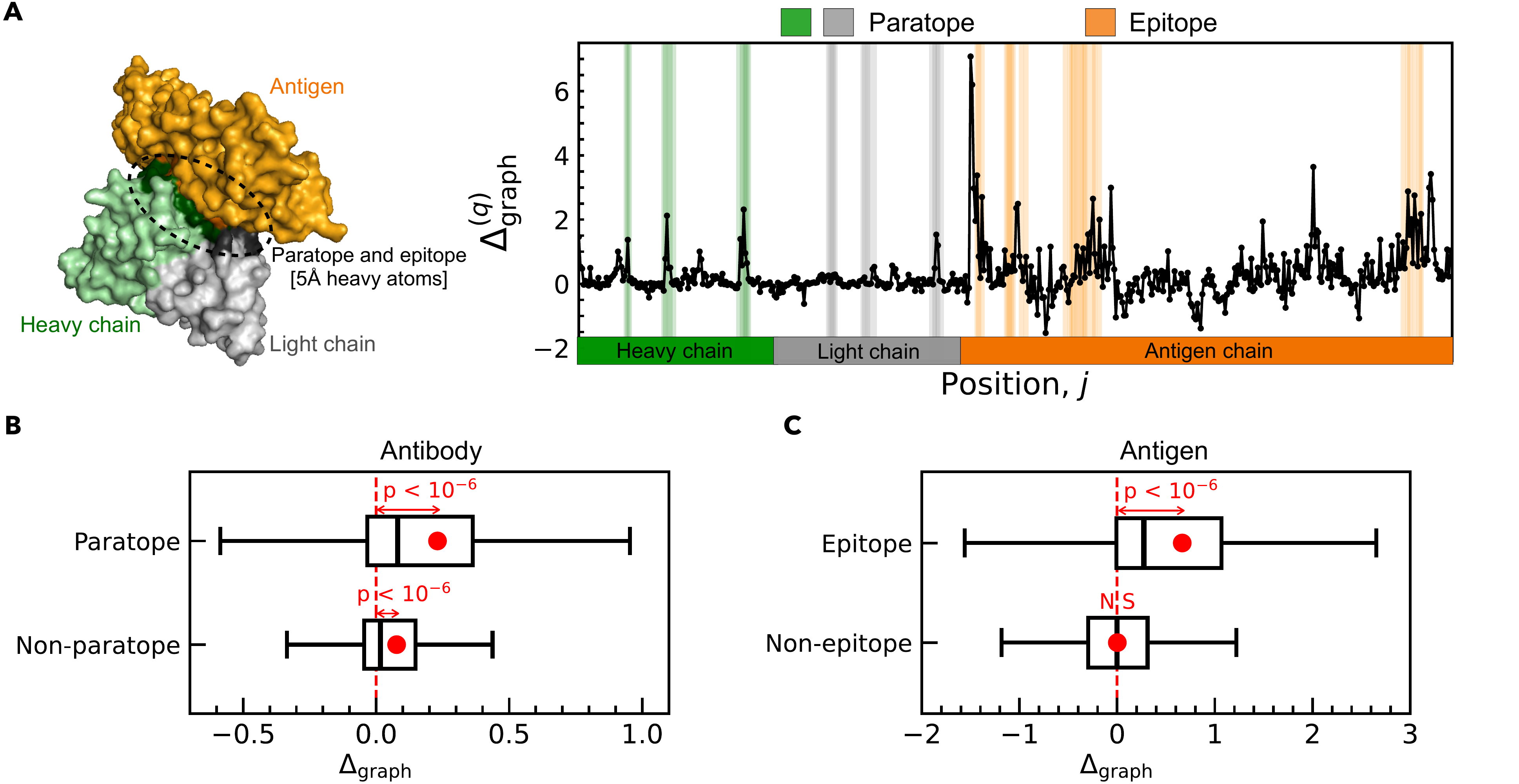}
\caption{\textbf{Graph-induced differential performance stratified by paratope-epitope membership}. (\textbf{A}) Left: Structure of the human monoclonal antibody 237235 in complex with \textit{Plasmodium vivax} reticulocyte binding protein 2b (PDB entry: \texttt{6wm9}), where we highlight in darker shade their interface, \textit{i.e.}, the paratope sites on the antibody and the epitope sites on the antigen, as defined by heavy atoms within $5\angstrom$. Right: $\Delta_{\mathrm{graph}}^{(q)}$ across residue positions $j$ for \texttt{6wm9}. (\textbf{B}) Boxplot of $\Delta_{\mathrm{graph}}$ for antibody paratope versus non-paratope members for the test set. (\textbf{C}) Boxplot of $\Delta_{\mathrm{graph}}$ for antigen epitope versus non-epitope members for the test set.}
\label{fig:4}
\end{figure}

Indeed, we systematically found across the entire test set that there is a significant graph-induced improvement for both paratope (higher) and non-paratope sites in the antibody but the improvement is significant only for epitope members on the antigen (\mfigrefs[B,C]{4}, \mtabrefs[E,F]{S2}), presumably because of the lower mean $\Delta_{\mathrm{graph}}^{(q)}$ of the antigen compared to the antibody (\figref[A,B]{3}) and because antigens proportionally contain more flexible regions like loops~\cite{Kunik2013}, for which $\Delta_{\mathrm{graph}}^{(q)}$ tends to be higher than, \textit{e.g.}, $\beta$-strands (\figref[B,C]{3}). This is supported by the INFUSSE and $S_{\mathrm{block}}$ errors shown in \mfigrefs[E,F]{S3}, which highlight the poor performance of the sequence-only model for paratope and epitope regions.


%
%
%
%
%

The improvement is significantly higher for paratope and epitope members compared, respectively, to non-paratope sites ($\Delta\mu=0.15$, p-value $<10^{-6}$) and non-epitope ($\Delta\mu=0.66$, p-value  $<10^{-6}$), see~\mfigrefs[B,C]{4} and~\mtabrefs[E,F]{S2}. Notably, almost the entire IQR lies on the positive side of the $\Delta_{\mathrm{graph}}$ axis for the paratope and the epitope, indicating that the benefit of graph-based modelling is broadly consistent. This enhancement is a result of the graph capturing the spatial constraints that effectively reduce the conformational flexibility of the paratope and epitope amino acids---a type of information not easily encoded in sequence statistics alone (see \mfigrefs[E,F]{S3}).

\begin{figure}[H]
\centering  \includegraphics[clip, trim=0cm 7.4cm 39cm 0cm, width=\linewidth]{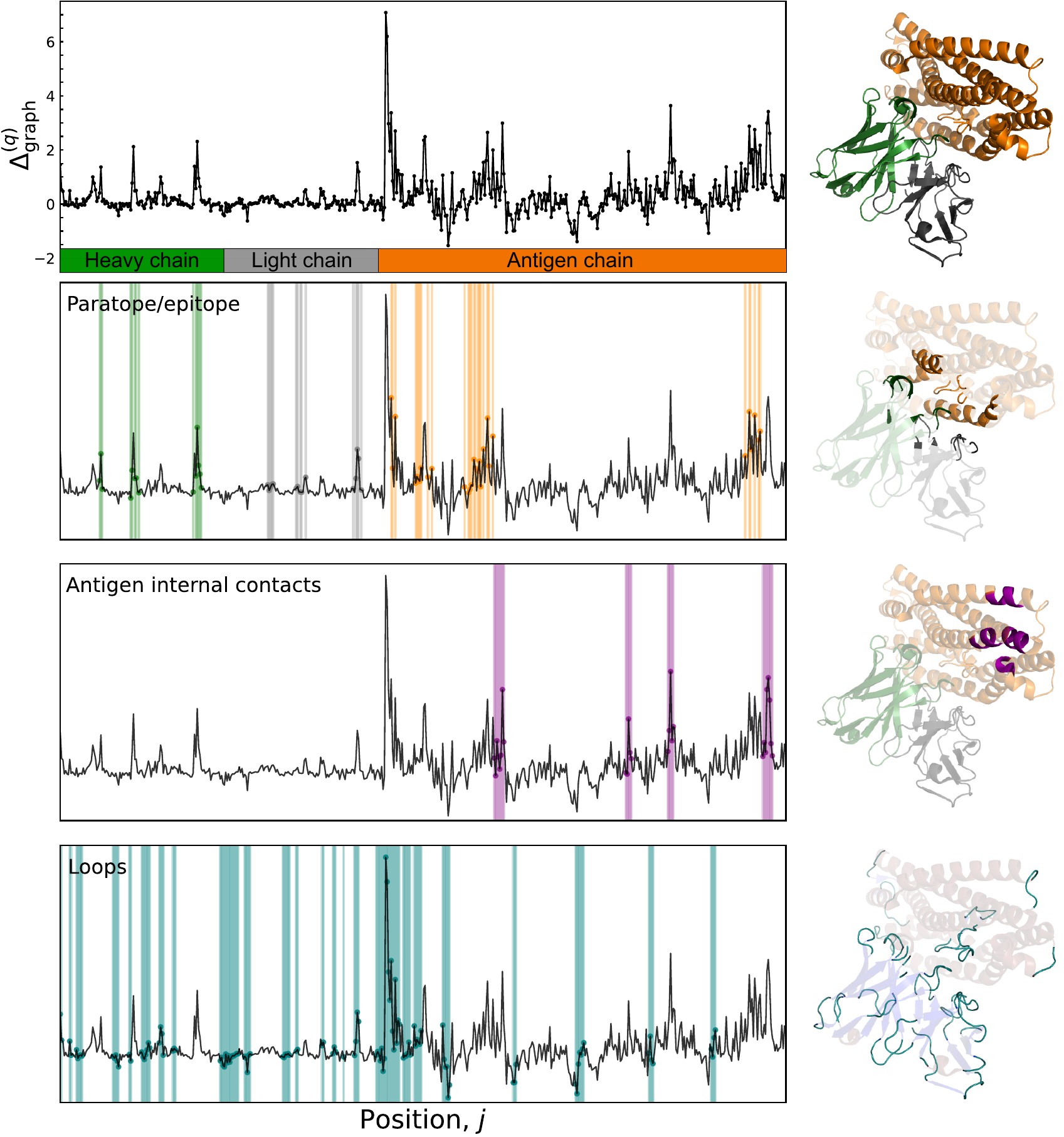}
\caption{\textbf{Detailed analysis of graph-induced differential performance for PDB entry \texttt{6wm9}.} $\Delta_{\mathrm{graph}}^{(q)}$ as a function of residue position $j$ for the PDB complex \texttt{6wm9} (first row) exhibits positive peaks corresponding to paratope and epitope members (second row, same plot as in Figure 4A) and internal contacts in the antigen (third row). Unstructured regions such as loops are associated to predominantly positive but also negative performance changes (fourth row).}
\label{fig:Fig5}
\end{figure}

Figure~\ref{fig:Fig5} shows an example where the values of $\Delta_{\mathrm{graph}}$ can be related to regions of high amino acid variability, loops and short $\alpha$-helices, and paratope/epitope members. We also found additional regions with large $\Delta_{\mathrm{graph}}$ that correspond to areas of internal, non-local interactions within the antigen linked to the global emergence of the tertiary structure of the protein. While $\Delta_{\mathrm{graph}}$ predominantly exhibits positive peaks for antigen internal contacts and binding site residues, the loop regions present a mixture of large positive peaks and some negative regions (mostly outside of the binding site). This in line with the wider IQR values found for these motifs, and consistent with the positive improvement of $\Delta_{\mathrm{graph}}$ over the whole test set.

\section{Discussion}

In this work, we have developed INFUSSE, a deep learning model for the prediction of single-residue properties in proteins that leverages sequence representations, both one-hot encodings and enriched amino acid embeddings learnt starting from a pre-trained LLM (ProtBERT), and integrates them with a structural graph using a diffusive Graph Convolutional Network (diff-GCN). We show that INFUSSE achieves greater accuracy for the prediction of antibody-antigen B-factors than existing methods, which typically neglect some of these types of information (Table 1). Within the graph block (diff-GCN), we used geometric graphs, having tested that detailed physicochemical graphs~\cite{Song2021} do not provide substantial advantage for this particular task of B-factor prediction. However, such constructions could prove useful for the prediction of other single-residue properties where the specific chemical structure, rather than molecular packing, plays a more significant role.

The block-modular structure (and training) of INFUSSE allowed us to generate insights into the predictability of a residue-level property, such as B-factors, from sequence statistics alone \emph{vs} with the addition of structural information on its molecular neighbourhood. Specifically, INFUSSE can be used to identify where in a given antibody-antigen complex the inclusion of a structural graph leads to improved residue-level predictions, as signalled by the differential prediction metric $\Delta_{\mathrm{graph}}$. 

In general, our study revealed a statistical association of  graph-induced predictive improvement 
with unstructured regions, positions with high residue variability, as well as paratope/epitope members. High composition variability and structural flexibility are main characteristics of the CDRs in antibodies, and are key properties to the maturation of high-affinity and specificity antigen binding. These regions exhibit also a distribution of $\Delta_{\mathrm{graph}}$ with wide interquartile ranges, reflecting the heterogeneity of the predictions derived from the sequence-only model.
%
%

These trends suggest that both the variability in sequence composition and the variability of possible conformations in unstructured and flexible regions pose challenges to sequence-only methods based on LLMs, such as ProtBERT, even after fine-tuning. 
Clearly, variability generally makes it practically difficult to train models with high generalisation power, due to the inevitable sparsity of the available sampling. More fundamentally, however, the properties of conformations in unstructured regions like loops are determined less strictly by the statistical patterns of sequence composition alone, in contrast to, \textit{e.g.}, $\beta$-sheets, where specific amino acid patterns translate into patterns of chemical bonds that underpin the stability of this structural motif. Rather, unstructured and flexible regions are sensitive to the molecular environment and the presence of binding partners. This lack of structural motif regularity implies weaker constraints acting less consistently to shape the sequence statistics across the large ensembles of protein sequences used as training datasets, which reduces the predictive power of sequence-based learning in highly diverse or unstructured regions, such as CDR loops~\cite{Ruffolo2023}. 
These considerations also explain the enhanced graph-induced predictive improvement for CDR loops and $\alpha$-helices observed here and can be seen in the disparity of performance of the sequence-only model in such regions.

This work opens several lines for future research. 
The prediction of B-factors with INFUSSE could be leveraged to better understand the role of local residue flexibility as a possible determinant of high binding affinity and specificity for antibodies. Specifically, characterising the correlates of high and low B-factors in antibody-antigen interfaces could provide insights into how affinity maturation is related to a gain and loss of rigidity of the CDR loops~\cite{Ovchinnikov2018}. 
Furthermore, analysing the computed graph-induced predictive improvement could provide hints towards principles for antibody design and protein engineering: high $\Delta_{\mathrm{graph}}$ suggests sites where amino acid composition and conformation are less subject to strong constraints, which could be targets for site-specific mutagenesis in the laboratory without compromising molecular viability. 
%
%

Another area of future work could exploit our results to inform how to refine the graph construction itself. Regions with low $\Delta_{\mathrm{graph}}$ values, such as $\beta$-strands, may not require dense connectivity, as the sequence block already performs well there. In contrast, motifs with large $|\Delta_{\mathrm{graph}}|$, such as intrinsically disordered or binding regions, might benefit from additional or longer-range graph edges. This opens the door to data-driven biomolecular graph construction strategies, where edge selection or weighting could be based on features like sequence entropy, structural prediction confidence or solvent exposure.
Similarly, the wide IQRs of $\Delta_{\mathrm{graph}}$ for loops suggest also that a single PDB snapshot, which only captures one conformation of an intrinsically disordered region, could be enhanced by using 
graph ensembles, \textit{e.g.}, from multiple molecular dynamics frames or nuclear magnetic resonance models. This approach would allow  learning over a set of occupancy-weighted graphs could provide a more faithful description of loop flexibility and further enhance the observed graph-induced improvement.

Finally, we remark that although we have focused, as an initial application, on the study of B-factors in antibodies, INFUSSE has been designed as a versatile node-specific regression model that could be straightforwardly used for other tasks, including epitope prediction or cysteine reactivity prediction~\cite{Marino2010}. This will be the object of future work.

\section{Methods}

\subsection{Data}
\label{sec:data}

Data were downloaded from the Structural Antibody Database (SAbDab)~\cite{SAbDab2014,SAbDab2022} on 4 October 2024 in the PDB format using Chothia numbering~\cite{Chothia1987}. Our download included all human and mouse antigen-bound and unbound antibody structures where paired heavy and light chains were present. We discarded structures with a resolution worse than $2.5 \angstrom$~\cite{ruffolo_antibody_2022}; structures without B-factors; and structures where the B-factor of any atom was higher than $100\angstrom^2$, as such high values have been linked to experimental errors~\cite{Carugo2018}. 
The resulting dataset contained $1510$ PDB entries for our analysis. 

\subsection{Construction of coarse-grained geometric protein graphs}
\label{sec:protein_graphs}

Each protein structure is represented as an undirected graph $\mathcal{G}$, 
where each of the $N$ nodes represents an amino acid and
$e_{ij}$, the edge connecting nodes $v_i$ and $v_j$, has weight $w(e_{ij})$ (for unweighted graphs, $w(e_{ij})=1$ if there is an edge $e_{ij}$ and $w(e_{ij})=0$ otherwise). 
The adjacency matrix $\mathbf{A}\in\mathbb{R}^{N\times N}$ of $\mathcal{G}$ is symmetric and its entries are given by $A_{ij} =  w(e_{ij})$.
Another matrix associated with $\mathcal{G}$ is the $N \times N$  Laplacian matrix $\mathbf{L}:=\mathbf{D}-\mathbf{A}$, where $\mathbf{D} = \text{diag}(\mathbf{A} \, \mathbf{1})\in\mathbb{R}^{N\times N}$ and $\mathbf{1}$ is the $N \times 1$ vector of ones (\textit{i.e.}, $\mathbf{D}$ contains the node degrees on the diagonal).  
For undirected graphs, $\mathbf{L}$ is symmetric 
and positive semi-definite~\cite{Stehlík2017}, with real non-negative eigenvalues and orthogonal eigenvectors. The graph Laplacian matrix is a central concept in spectral graph theory and is directly linked to diffusive processes on graphs~\cite{lambiotte2014}.

We consider two geometric graph constructions. Let $\mathbf{r}_i, \mathbf{r}_j \in \mathbb{R}^3$ be the positions of the $\alpha$-carbon atoms of amino acids $i$ and $j$, respectively. Our main construction is a simple \textit{weighted Gaussian graph} with (full) adjacency matrix given by:
\begin{align}
 A_{ij}=\exp\left(-\frac{\lVert\mathbf{r}_i - \mathbf{r}_j\rVert_2^2}{\eta^2}\right)  
 \label{eq:Gaussiangraph}
\end{align}
where $|| \cdot ||_2$ stands for the Euclidean distance and  $\eta=8$ as in Ref.~\cite{Opron2016}.
For comparison, we also consider (unweighted) Gaussian Network Models (GNMs), based on an $\epsilon$-ball cutoff~\cite{Liu2020}, whereby:
\begin{align}
   A_{ij} =
    \begin{cases}
      1, & \text{ if } \lVert \mathbf{r}_i -  \mathbf{r}_j \rVert_2 < \epsilon\\
      0, & \text{ otherwise}
    \end{cases}      
    \label{eq:GNM}
\end{align}
with a cutoff distance of $\epsilon=8\angstrom$ and $\epsilon=10\angstrom$ as in Ref.~\cite{Baldi2003}.




\subsection{LLM sequence embeddings}
\label{sec:llm_embeddings}

As part of our pipeline, we generate embeddings from a \textit{frozen} ProtBERT model~\cite{Elnaggar2022}, a $30$-layer Bidirectional Encoder Representations from Transformers (BERT) LLM pre-trained on 217 million protein sequences from UniRef100~\cite{Suzek2015}. The input to ProtBERT is tokenised protein sequences comprising the $20$ standard amino acids, with additional tokens for rare amino acids, chain start and chain separation. In this work, the protein sequences we consider are concatenations of light and heavy antibody chains and the antigen (for antigen-bound antibody structures), or of light and heavy chain for unbound structures (see Figure \ref{fig:gcn-bf}). For each tokenised protein sequence $\mathbf{x}_{\mathrm{s}}$ with $N$ amino acids, we obtain an embedding given by the output layer of ProtBERT, $\mathbf{E}_{\mathrm{s}}\in\mathbb{R}^{N\times F_{\mathrm{E}}}$ with embedding dimension $F_{\mathrm{E}}=1024$. 

\subsection{The diffusive GCN architecture}
\label{sec:diffGCN}

\paragraph*{GCN background and notation.} Graph Convolutional Networks (GCNs)~\cite{Kipf2017} are deep learning architectures that use a layer propagation rule derived from a first-order approximation of spectral graph convolutions.
Consider a graph $\mathcal{G}$ with Laplacian $\mathbf{L}$, adjacency matrix $\mathbf{A}$, and normalised Laplacian $\mathbf{\widetilde{L}}:= \mathbf{D}^{-\frac12} \mathbf{L} \mathbf{D}^{-\frac12} = \mathbf{I}-\mathbf{D}^{-\frac12}\mathbf{A}\mathbf{D}^{-
\frac12}$ with eigendecomposition  $\mathbf{\widetilde{L}}=\mathbf{U}\mathbf{\Lambda}\mathbf{U}^T$.

If $\mathbf{x} \in \mathbb{R}^N$ is a signal defined on the $N$ nodes, then $\mathbf{\hat{x}} = \mathbf{U}^T \mathbf{x}$ is its graph Fourier transform. 
Let $\mathbf{g}_{\bm{\theta}} = \text{diag}(\bm{\theta})$ be a filter in the Fourier domain, where $\bm{\theta} \in \mathbb{R}^N$ is a vector of parameters. A spectral convolution on $\mathcal{G}$ is then given by:
\begin{align*}
\mathbf{y}= \mathbf{U}\mathbf{g}_{\bm{\theta}}\mathbf{\hat{x}} = \mathbf{U}\mathbf{g}_{\bm{\theta}}\mathbf{U}^T\mathbf{x}    
\end{align*}
which highlights that $\mathbf{g}_{\bm{\theta}}$ is a function of the eigenvalues of $\mathbf{\widetilde{L}}$. 
To avoid the costly eigendecomposition of $\mathbf{\widetilde{L}}$,  $\mathbf{y}$ is approximated in terms of Chebyshev polynomials~\cite{Hammond2011}:
\begin{align*}
\mathbf{y}\approx \sum_{k=0}^{K-1}\theta'_k\mathbf{T}_k(\mathbf{L}_{\mathrm{scaled}})\mathbf{x}    
\end{align*}
where $\bm{\theta'}\in\mathbb{R}^K$ are Chevyshev coefficients; 
$\mathbf{ L}_{\mathrm{scaled} }:=\frac{2}{\lambda_{\text{max}}}\mathbf{\widetilde{L}}-\mathbf{I}$ is the scaled Laplacian with $\lambda_{\text{max}}$ being the largest eigenvalue of $\mathbf{\widetilde{L}}$;
and $\mathbf{T}_k(\mathbf{L }_{\mathrm{scaled}})\in \mathbb{R}^{N\times N}$ is the $k$-th order Chebyshev polynomial evaluated for $\mathbf{L }_{\mathrm{scaled}}$.
To first-order ($K=1$, $\theta'_0=2$, $\theta'_1=-1$), the output $\mathbf{y}$ simplifies to:
\begin{align*}
\mathbf{y}\approx (\mathbf{I}+\mathbf{D}^{-\frac12}\mathbf{A}\mathbf{D}^{-\frac12})\mathbf{x} =
( 2 \, \mathbf{I} - \mathbf{\widetilde{L}}) \mathbf{x}
\end{align*}
Numerical instability in this expression is avoided through a renormalisation trick (akin to adding self-loops to the graph) via the substitution: 
$(\mathbf{I} + \mathbf{D}^{-\frac{1}{2}} \mathbf{A} \mathbf{D}^{-\frac{1}{2}}) 
\mapsto 
(\mathbf{\widetilde{D}}^{-\frac{1}{2}} \mathbf{\widetilde{A}} \mathbf{\widetilde{D}}^{-\frac{1}{2}})$, where $\mathbf{\widetilde{A}} = \mathbf{A} + \mathbf{I}$ and $\mathbf{\widetilde{D}} = \text{diag} (\mathbf{\widetilde{A}} \, \mathbf{1} )$. This approximation of the spectral graph convolution is applied at every layer of the GCN, as follows.
Let $\mathbf{H}^{(l)} \in \mathbb{R}^{N \times F_l}$ denote the input signal for $N$ nodes and $F_l$ channels at the $l^{\text{th}}$ layer ($l=0,\dots,L-1$).
Then the output of the GCN $l^{\text{th}}$ layer is given by:
\begin{align}
    \mathbf{H}^{(l+1)}:=\sigma\left(
    (\mathbf{\widetilde{D}}^{-\frac12}\mathbf{\widetilde{A}}\mathbf{\widetilde{D}}^{-\frac12}  ) \, \mathbf{H}^{(l)}\mathbf{W}^{(l)}\right),
    \label{eq:GCN}
\end{align}
where $\mathbf{W}^{(l)} \in \mathbb{R}^{F_{l+1} \times F_l}$ is the matrix of learnable weights for the $l^{\text{th}}$ layer, and $\sigma$ is a non-linearity. 

\paragraph*{The diffusive GCN for protein graphs and sequence embeddings.}
Although GCNs were originally designed for node classification, they can be adapted for regression, as in our task here. In this work, the input $\mathbf{H}^{(0)}$ corresponds to the node embeddings $\mathbf{X} \in \mathbb{R}^{N \times F}$ learnt starting from the ProtBERT embeddings $\mathbf{E}_{\mathrm{s}}\in\mathbb{R}^{N\times F_{\mathrm{E}}}$ of the input protein sequence (see Eq. \eqref{eq:Sblock}). To avoid over-smoothing~\cite{Zhou2023, ChenLin2020}, our GCN architecture consists of two layers: a first layer with a ReLU non-linearity, and a second layer without non-linearity. The number of internal units between the first and second layer is kept at $F$. Furthermore, as discussed in Ref.~\cite{Peach2020}, the renormalised operator $(\mathbf{\widetilde{D}}^{-\frac{1}{2}} \mathbf{\widetilde{A}} \mathbf{\widetilde{D}}^{-\frac{1}{2}})$ is replaced by a graph diffusion operator to enhance learning through the graph. This gives us our \textit{diffusive Graph Convolutional Network} (diff-GCN) model, whose output $\mathbf{H}^{(2)}$ is given by:
\begin{align}
\text{diff-GCN}_t(\mathbf{L},\mathbf{X}):= \mathbf{H}^{(2)} = e^{-t \mathbf{ L}} \, \text{ReLU}\left(e^{-t \mathbf{ L}} \mathbf{X} \, \mathbf{W}^{(0)}\right) \mathbf{W}^{(1)},
\label{eq:diffGCN}
\end{align}
where $e^{-t \, \mathbf{L}}$ is the diffusion transition matrix associated with the graph Laplacian $\mathbf{L}$, and $t$ is a learnable scale parameter. 
Note that the learnable weight matrices $\mathbf{W}^{(0)} \in \mathbb{R}^{F \times F}$ and $\mathbf{W}^{(1)} \in \mathbb{R}^{F \times 1}$ are independent of the number of amino acids, $N$. Therefore, the diff-GCN learns a mapping between input node embeddings and predicted node properties without the need for sequence alignment. 

\subsection{Full INFUSSE architecture}
\label{sec:full_infusse}
The full INFUSSE architecture is composed of a sequence block and a graph block, as follows.  

\paragraph*{Sequence block:} The one-hot encoded representation $\mathbf{\Phi}(\mathbf{x}_{\mathrm{s}})\in\lbrace{0,1\rbrace}^{N\times d}\subset\mathbb{R}^{N\times d}$ of a sequence of an antibody-antigen complex $\mathbf{x}_{\mathrm{s}} \in\mathbb{N}^N$ and the ProtBERT output layer embedding $\mathbf{E}_{\mathrm{s}}:=\mathrm{ProtBERT(\mathbf{x}_{\mathrm{s}})}\in\mathbb{R}^{N\times F_{\mathrm{E}}}$, with $F_E=1024$, are combined via learnable transformations $T_1:\mathbb{R}^{N\times d}\rightarrow\mathbb{R}^{N\times F}$ and $T_2:\mathbb{R}^{N\times F_{\mathrm{E}}}\rightarrow\mathbb{R}^{N\times F}$, respectively, which are summed to give an enriched sequence representation $\mathbf{X} \left(\mathbf{x}_{\mathrm{s}}, \mathbf{E}_{\mathrm{s}} \right)$ that leverages the patterns learnt by the LLM (embedded by $\mathbf{E}_{\mathrm{s}}$) but is further fine-tuned on the specific protein data and prediction task of interest via $T_1$ and $T_2$. $\mathbf{X}$ is next passed as input to an additional transformation $T_3:\mathbb{R}^{N\times F}\rightarrow\mathbb{R}^{N}$, where the embedding dimension $F$ is a hyperparameter.
All three transformations $T_1, T_2, T_3$ are implemented as linear layers followed by non-linear (ReLU) activation functions, \textit{e.g.}, $T_2 (\mathbf{E}_{\mathrm{s}}) = \text{ReLU} (\mathbf{E}_{\mathrm{s}} \, \bm{\mathcal{W}}_2)$ where $\bm{\mathcal{W}}_2 \in \mathbb{R}^{F_{\mathrm{E}}\times F}$ is learnable, 
and similarly for $T_1$ and $T_3$. Here, $d=21$ as it accounts for the $20$ standard residues and rare amino acids. The output of the \textit{sequence block} is thus given by:
\begin{align}
    S_{\mathrm{block}} \left(\mathbf{x}_{\mathrm{s}}, \mathbf{E}_{\mathrm{s}} \right) := T_3 \left( \mathbf{X}   \right) 
    \quad \text{with} \quad
    \mathbf{X} \left(\mathbf{x}_{\mathrm{s}}, \mathbf{E}_{\mathrm{s}} \right) :=T_1\left(\mathbf{\Phi}(\mathbf{x}_{\mathrm{s}})\right) +  T_2(\mathbf{E}_{\mathrm{s}}). 
    \label{eq:Sblock}
\end{align}
 
\paragraph*{Graph block:} From the spatial coordinates $\mathbf{r}\in\mathbb{R}^{3N}$ of the $N$ $\alpha$-carbons of the protein complex we  build a geometric graph with adjacency matrix $\mathbf{A}(\mathbf{r})$ (\textit{e.g.}, as in Eq.~\eqref{eq:Gaussiangraph}) and Laplacian $\mathbf{L}(\mathbf{r}) \in \mathbb{R}^{N\times N}$. 
This graph Laplacian and the enriched sequence representations $\mathbf{X}\left(\mathbf{x}_{\mathrm{s}}, \mathbf{E}_{\mathrm{s}} \right)
\in\mathbb{R}^{N\times F}$ obtained as part of the sequence block in Eq.~\eqref{eq:Sblock} are the inputs for the diff-GCN in Eq.~\eqref{eq:diffGCN}.
The \textit{graph block} is then:
\begin{align}
   G_{\mathrm{block}}(\mathbf{r}, \mathbf{X}):=\text{diff-GCN}_t(\mathbf{L}(\mathbf{r}),\mathbf{X}
   ), 
   \label{eq:Gblock}
\end{align}
with the scale parameter $t$ as a learnable parameter. 

\paragraph*{INFUSSE model:} The outputs of the sequence and graph blocks are summed to give the predicted single-residue properties (B-factors). 
The full INFUSSE model (see Figure~\ref{fig:gcn-bf}) is then given by:
\begin{equation}
\text{INFUSSE}(\mathbf{x}_{\mathrm{s}}, \mathbf{r}):= S_{\mathrm{block}} \left(\mathbf{x}_{\mathrm{s}}, \mathbf{E}_{\mathrm{s}} \right) +
G_{\mathrm{block}}(\mathbf{r}, \mathbf{X} (\mathbf{x}_{\mathrm{s}}, \mathbf{E}_{\mathrm{s}})
)
\end{equation}
\subsection{Training of INFUSSE}
\label{sec:training}
To avoid redundancy between the training and test sets, we choose an antibody-antigen complex at random and add it to the test set only if all its chains have less than 90\% sequence identity to any remaining corresponding chain in the training set; otherwise, it is rejected and reassigned to the training set~\cite{Davila2022}. This process is repeated until the desired size of the test set is reached, see Ref.~\cite{Michalewicz2024}. We produced 10 such splits to test the robustness of our results.

As described above, the training proceeds in two steps. In \textit{step 1}, we train the sequence block $S_{\mathrm{block}}$ (Eq.~\eqref{eq:Sblock}) to predict B-factors from sequence information $\mathbf{x}_{\mathrm{s}}$ and the LLM (ProtBERT) output $\mathbf{E}_{\mathrm{s}}$; this is effectively equivalent to fine-tuning the ProtBERT embeddings together with one-hot encodings for this specific downstream task by learning $T_1, T_2, T_3$. 
In \textit{step 2}, the weights for $S_{\mathrm{block}}$ obtained in \textit{step 1} are used as initialisation, and we incorporate the graph block (Eq.~\eqref{eq:Gblock}) to optimise the transformations $T_1, T_2, T_3$ jointly with the learnable parameters of the graph block (the weight matrices $\mathbf{W}^{(0)}, \mathbf{W}^{(1)}$ and the scale parameter $t$, see section~\ref{sec:diffGCN}).

Specifically, let $\mathcal{P}$ be the training set of antibody-antigen complexes and unbound antibodies from PDB, where each PDB entry $p \in \mathcal{P}$, with $N_p$ residues, contains the sequence $\mathbf{x}^{(p)}_{\mathrm{s}}$, the $\alpha$-carbon 3D coordinates $\mathbf{r}^{(p)}$, and a vector of ground truth standardised B-factors $\mathbf{B}^{(p)}=(B^{(p)}_1, \ldots, B^{(p)}_{N_p})$. Since we are performing a regression task, training proceeds by minimising the mean squared error between the ground truth and predicted B-factors over the training set:  
\begin{align}
\left \langle  \frac{1}{N_p} \left \lVert \mathbf{B}^{(p)} - \hat{\mathbf{B}}^{(p)} 
\right \rVert_2^2 \right \rangle_{p \in \mathcal{P}}, 
\label{eq:MSE_total}
\end{align}
where $\langle \cdot \rangle_{p \in \mathcal{P}}$ denotes the average over the $|\mathcal{P}| = 1435$ entries in the training set, and $\hat{\mathbf{B}}^{(p)}$ denotes the vector of predicted B-factors obtained from INFUSSE during the two steps:
\begin{align}
\text{\textit{Step 1:}} \quad \hat{\mathbf{B}}^{(p)} &=  S_{\mathrm{block}} (\mathbf{x}^{(p)}_{\mathrm{s}}, 
 \mathbf{E}^{(p)}_{\mathrm{s}} ) 
 \label{eq:B_Sblock}\\
 \text{\textit{Step 2:}} \quad \hat{\mathbf{B}}^{(p)} & = 
 \text{INFUSSE}(\mathbf{x}^{(p)}_{\mathrm{s}}, \mathbf{r}^{(p)}).
 \label{eq:B_infusse}
\end{align}
where the \textit{step 2} predictions by INFUSSE combine the structure and graph blocks (Eq.~\ref{eq:full_infuse}). In contrast, during \textit{step 1} the graph block is deactivated, hence the predictions $\hat{\mathbf{B}}^{(p)}$ are given only by $S_{\mathrm{block}}$.

The minimisation is carried out using the AdamW optimiser with learning rate $l_r=3\times10^{-3}$. AdamW decouples weight decay from the gradient update process, leading to faster convergence and better generalisation~\cite{Zhou2023AdamW}, and incorporates $\ell_2$ regularisation with a Lagrange parameter $\lambda=10^{-2}$ to avoid overfitting. The optimal combination of learning rate and embedding size was found through a grid search using $10$-fold cross-validation, and yielded $l_r=3\times10^{-3}$ and $F=16$. We considered all combinations of learning rates in $\lbrace 10^{-5}, 10^{-4}, 5\times 10^{-4}, 10^{-3}, 5\times 10^{-3}, 3\times 10^{-3}, 2\times 10^{-3}, 10^{-3}, 5\times 10^{-2}\rbrace$ and embedding sizes in $\lbrace 4, 8, 16, 32, 64, 128\rbrace$, and chose the combination with the lowest validation mean squared error across the 10 folds. The training was done using one NVIDIA GeForce RTX 3090 Ti~GPU.

\subsection{Model evaluation and post-hoc interpretation}
\label{sec:methods_evaluation}
We evaluated the performance of INFUSSE and of alternative modelling approaches (Table~\ref{table:gcn-bf-perf} and S1) on each test set $\mathcal{Q}$ containing $|\mathcal{Q}| = 75$ complexes as follows. For each PDB entry $q \in \mathcal{Q}$, with $N_q$ residues, we obtain the predicted INFUSSE B-factors  
$\hat{B}^{(q)}_j = [\text{INFUSSE}(\mathbf{x}^{(q)}_{\mathrm{s}}, \mathbf{r}^{(q)})]_j, \, j=1, \ldots, N_q$, 
and compute the Pearson correlation coefficient $R$ for the set of paired predicted and true B-factors in complex $q$ $\{ \hat{B}^{(q)}_j,B^{(q)}_j  \}_{j=1}^{N_q}$. We then  average over all the complexes in the test set to obtain the performance for the corresponding training-test split:
\begin{align*}
R_\mathcal{Q} = \left \langle R^{(q)} \right \rangle_{q \in \mathcal{Q}} 
\quad \text{where} \quad 
%
R^{(q)} = \text{PCC} \left( \{ \hat{B}^{(q)}_j,B^{(q)}_j   \}_{j=1}^{N_q} \right ), \, 
q=1, \dots, |\mathcal{Q}| 
\end{align*}
This process is repeated for five different training/test splits (with different random seeds) to obtain the average $R$
across the different splits.

\subsection{Computation of the graph-induced differential performance} 
\label{sec:delta_graph}
To quantify the difference introduced by the full INFUSSE model relative to the LLM-based only-sequence $S_{\mathrm{block}}$ model, we compute the prediction errors of both models for the $j=1, \ldots, N_q$ residues of each of the PDB entries $q \in \mathcal{Q}$ in the test set, given respectively by:
\begin{align*}
\varepsilon^{(q)}_{S_{\mathrm{block}}, j} &:= \left(\hat{B}^{(q)}_{S_{\mathrm{block}},j} - B^{(q)}_j \right)^2 = 
\left([
    S_{\mathrm{block}} (\mathbf{x}^{(q)}_{\mathrm{s}}, 
 \mathbf{E}^{(q)}_{\mathrm{s}})]_j -B^{(q)}_j \right)^2,\\
    \varepsilon^{(q)}_{\text{INFUSSE}, j} &:= \left(\hat{B}^{(q)}_{j} - B^{(q)}_j \right)^2 = \left( [\text{INFUSSE}(\mathbf{x}^{(q)}_{\mathrm{s}}, \mathbf{r}^{(q)})]_j -B^{(q)}_j \right)^2 
\end{align*}
and we collect the differences in prediction errors for all residues $j$ for each complex $q$ into a set:
\begin{align*}
\Delta_{\mathrm{graph}} =
\left \{  
    \Delta^{(q)}_{\text{graph},j}
    \right \} \quad \text{where} \quad 
   \Delta^{(q)}_{\text{graph}, j} &:= 
   \varepsilon^{(q)}_{S_{\mathrm{block}}, j}-\varepsilon^{(q)}_{\text{INFUSSE}, j}    \quad j =1,\cdots, N_q,\, q=1, \cdots, |\mathcal{Q}| .
   \label{eq:Delta_fullset}
\end{align*}

A positive value of $\Delta^{(q)}_{\text{graph},j}$ indicates that adding the structural
graph \textit{reduces} the squared error for residue $j$ in complex~$q$, whereas
a negative value means that the graph increases the squared error. For
each biological scenario described below, we use the full set $\Delta_{\mathrm{graph}}$ to assess via non-parametric bootstrap tests (see Methods, Section \ref{sec:statistical_tests}) whether: (i) the group mean of $\Delta_{\mathrm{graph}}$ is significantly greater than zero, (ii) whether the group means of two scenarios are significantly different, and  (iii) whether the interquartile ranges (IQRs) of two scenarios are significantly different.

\subsection{Biological and structural scenarios }
\label{sec:scenarios}
We employ statistical tests to compare between groups informed by the following biological scenarios. 
\begin{itemize} 
\item \textbf{Sequence variability and diversity scores: }
We aligned the antibody sequences using Antibody Numbering and Antigen Receptor ClassIfication (ANARCI)~\cite{ANARCI} under the Chothia convention, to ensure consistent positions for the antibody residues belonging to same regions across the dataset. For each antibody, we only keep the variable region, as there is no standard alignment convention for the constant region. 

Let $c(n,aa)$ be the count of residue type $aa$ ($aa= 1,\dots, 20$) in position $n$ ($n=1,\dots, N^*$) in the training set, where 
$N^*$ represents the maximum position index across all sequences after ANARCI alignment. The residue probabilities at each position are then given by:
\begin{align*}
p(n, aa) := \frac{c(n,aa)}{\sum_{aa} c(n,aa)}, 
\end{align*}
and the Shannon entropy for each position $n$ is:
\begin{align*}
    \mathcal{H}(n) := -\sum_{aa} p(n, aa)\log_2{p(n, aa)}.
\end{align*}
For each PDB entry $q$ in the test set $\mathcal{Q}$ with $N_q$ residues and sequence $\mathbf{x}^{(q)}_s$, we compute diversity scores $D(\mathbf{x}^{(q)}_s)$ as the set of entropies for all the positions $n \in \{ n_1,\dots,n_{N_q} \}$, where $\{ n_1,\dots,n_{N_q} \}$ denotes the subset of positions in the aligned version of sequence $q$ of its $N_q$ amino acids. We then compile them into a set containing the scores for all the complexes in the test set:
\begin{equation}
\left \{  D(\mathbf{x}^{(q)}_s) \right \}_{q=1}^{|\mathcal{Q}|}, \quad \text{where} \quad
D(\mathbf{x}^{(q)}_s) := \left \{ \mathcal{H}(n): 
n \in \{ n_1,\dots,n_{N_q} \}
\label{eq:diversity_scores_set}
\right \}
\end{equation}
We then distribute the diversity scores of the test data into high- and low-entropy groups (in bits), each containing half of the data points, such that the separating value (median) is $1.457$ $\text{bits}$.

\item \textbf{Structural motifs: } Since PDB files categorise each amino acid as part of $\alpha$-helices, $\beta$-sheets or loops, we use these labels to test the association of secondary structure in antibody and antigens with large values within the set $\Delta_{\mathrm{graph}}$.

\item \textbf{Paratope-epitope sites: } A paratope member is an antibody residue such that any of its heavy atoms is within a distance of less than $5\angstrom$ from any heavy atom of an amino acid in the antigen, and conversely for epitope members~\cite{Akbar2021}. For each antibody-antigen complex, we identify residues that belong to the paratope/epitope binding interface, and we analyse if high $\Delta_{\mathrm{graph}}$ values are more prevalent in those regions.
\end{itemize}

\subsection{Statistical significance tests}
\label{sec:statistical_tests}
To assess whether the distribution of the graph-induced differential performance $\Delta_{\mathrm{graph}}$ has significant differences in relation to particular structural or biological properties, we use non-parametric statistical tests as follows (see also main text and Methods, Section \ref{sec:scenarios}). 

Let each residue $j$ in the protein complex $q$ of the test set $\mathcal{Q}$ have a binary label $\mathrm{c}_j^{(q)}\in\{0,1\},\, j=1,\dots,N_q,\,q=1,\dots,|\mathcal{Q}|$ reflecting data stratification into two groups according to a particular property. For instance, the labels $\mathrm{c}_j^{(q)}$ could represent whether an amino acid belongs to a framework (`0') or CDR (`1'). Recall also that the set $\Delta_{\mathrm{graph}} =
\left \{  
    \Delta^{(q)}_{\text{graph},j}
    \right \} $ contains all the differential squared error predictions of this set of residues. 

Let $N_k$ be the number of amino acids in the test set with label $k$, and let us define the empirical mean of $\Delta^{(q)}_{\mathrm{graph}, j}$ for residues with label $k$:
\begin{align*}
\hat{\mu}_k:=\frac{1}{N_k}\sum_{q=1}^{|\mathcal{Q}|} \sum_{j=1}^{N_q} \Delta^{(q)}_{\mathrm{graph}, j}\mathbbm{1}_{\mathrm{c}_j^{(q)}=k},
\end{align*}
where $\mathbbm{1}_{\mathrm{c}_j^{(q)}=k}$ is the indicator function that equals $1$ if the $j^{\mathrm{th}}$ residue in the $q^{\mathrm{th}}$ sample has label $k$ and 0 otherwise.  

Specifically, our statistical tests attempt to reject that the following statistics are equal to zero: (i) the mean of $\Delta_{\mathrm{graph}}$ for each group; (ii) the difference  of means of $\Delta_{\mathrm{graph}}$ of the two groups; and (iii) the differences of interquartile ranges of $\Delta_{\mathrm{graph}}$ of the two groups.

For (ii), we formulate a hypothesis test to compare population means $\mu_0$ and $\mu_1$ of classes $k=0$ and $k=1$,
null hypothesis:
\[ H_0: \, \Delta\mu := \mu_1 - \mu_0 = 0. \]
We compute the observed statistic as $t_{\mathrm{obs}}:=\hat{\mu}_1-\hat{\mu}_0$ and we follow a bootstrap procedure with $S$ resamples, with $S=10^6$. For each resample, we create a bootstrap test set by sampling with replacement $N_0$ values of $\Delta^{(q)}_{\mathrm{graph}, j}$ with class $k=0$, and $N_1$ values of $\Delta^{(q)}_{\mathrm{graph}, j}$ with class $k=1$. We then compute $t^s:=\hat{\mu}_1^s-\hat{\mu}_0^s, s=1,\dots,S$. The p-value is then estimated as the fraction of bootstrap samples where $t^s \geq t_\mathrm{obs}$, corresponding to the achieved significance level (ASL)~\cite{Efron1994}:
\begin{align*}
  \widehat{\mathrm{ASL}}=\frac{\#\{{t^s\geq t_{\text{obs}}}\}}{S}  
\end{align*}

For (iii), we repeat the procedure as in (ii) but for the hypothesis 
\[H_0: \, \Delta\mathrm{IQR}:=\mathrm{IQR}_1-\mathrm{IQR}_0 = 0.
\]

For (i), we apply the same bootstrap procedure as in (ii) to reject the hypothesis that the mean of $\Delta^{(q)}_{\mathrm{graph}, j}$ for a group $k$ is equal to zero:
\[ H_0: \, \mu_k = 0. \]
This one-sample test allows us to conclude if adding graph information improves performance.

For cases with a number of labels $k>2$ (\figref[C]{2} and \mfigrefs[B,C]{3}), we assess statistical significance through pairwise tests. Significance tests (ii) and (iii) are performed also for the predictions of $S_{\mathrm{block}}$ and INFUSSE by replacing $\Delta_{\mathrm{graph}}$ with, respectively, $\varepsilon_{S_{\mathrm{block}}}$ and $\varepsilon_{\mathrm{INFUSSE}}$, evaluating the same categorical groupings.

\subsection*{Python package}

The code is available as a Python package in the GitHub repository~\url{github.com/kevinmicha/INFUSSE}, which includes installation and dependency management instructions and ready-to-run environments. The code is modular so that any pre-trained LLM (other than ProtBERT) or other manually constructed input node embeddings can be used, and any graph construction can be incorporated with ease. 

\section*{Acknowledgements}

K.M. acknowledges support from the President's PhD Scholarship at Imperial College London. M.B. acknowledges support by the Engineering and Physical Sciences Research Council (EPSRC) under grant EP/N014529/1 funding the EPSRC Centre for Mathematics of Precision Healthcare at Imperial College London.
All authors are grateful to Daniyar Ghani, Daniel Salnikov and Yinfei Yang for their valuable suggestions on statistical analyses, as well as to Alessia David for helpful discussions.

\section*{Author contributions}

Study concept and design: K.M., M.B., and B.B.; Development of source
code: K.M.; Analysis and interpretation of data: K.M., M.B., B.B.; Writing and revision of the manuscript: K.M., M.B., and B.B.;
Study supervision: M.B. and B.B.

\section*{Declaration of interests}

The authors declare no competing interests.

\bibliographystyle{unsrt}
\bibliography{references} 

\newpage
\section*{Supplemental information}
\markboth{Supplemental information}{Supplemental information}

\setcounter{figure}{0}
\setcounter{page}{1}
\setcounter{table}{0}
\renewcommand{\thepage}{S\arabic{page}}  
\renewcommand{\thesection}{S\arabic{section}}  
\renewcommand{\thetable}{S\arabic{table}}
\renewcommand{\thefigure}{S\arabic{figure}}

\begin{table}[h!]
\centering
\footnotesize
\caption{\textbf{Full summary of performance for different methods.}}
\resizebox{\textwidth}{!}{
\begin{tabular}{|l|c|c|c|c|}
\hline
\textbf{Method} & \textbf{Sequence representation} & \textbf{Structure representation} & \textbf{Learnt with ML} & $R$ \\ \hline
\multirow{3}{*}{INFUSSE ($S_{\mathrm{block}}$ + $G_{\mathrm{block}}$)} 
    & \multirow{3}{*}{One-hot encoding \& LLM embeddings} & Weighted Gaussian graph ($\eta=8$) & \multirow{3}{*}{$T_1, T_2, T_3, \mathbf{W}^{(l)}, t$} & $0.71$ \\ \cline{3-3} \cline{5-5}
    &  & GNM ($\epsilon=10\angstrom$) &  & $0.70$ \\ \cline{3-3} \cline{5-5}
    &  & GNM ($\epsilon=8\angstrom$) &  & $0.69$ \\ \hline
\multirow{3}{*}{$S_{\mathrm{block}}$ + $G_{\mathrm{block}}$ (GCN instead of diff-GCN)} 
    & \multirow{3}{*}{One-hot encoding \& LLM embeddings} & Weighted Gaussian graph ($\eta=8$) & \multirow{3}{*}{$T_1, T_2, T_3, \mathbf{W}^{(l)}$} & $0.69$ \\ \cline{3-3} \cline{5-5}
    &  & GNM ($\epsilon=10\angstrom$) &  & $0.69$ \\ \cline{3-3} \cline{5-5}
    &  & GNM ($\epsilon=8\angstrom$) &  & $0.68$ \\ \hline
$S_{\mathrm{block}}$ alone (no structure) & One-hot encoding \& LLM embeddings & --- & $T_1, T_2, T_3$ & $0.64$ \\ \hline
$S_{\mathrm{block}}$ (no LLM) + $G_{\mathrm{block}}$ & One-hot encoding & Weighted Gaussian graph ($\eta=8$) & $T_1, T_3, \mathbf{W}^{(l)}, t$ & $0.55$ \\ \hline \hline
LSTM~\cite{Pandey2023} (SOTA for general proteins) & One-hot encoding & Raw coordinates $\mathbf{r}$, secondary structure, and chain breaks  & LSTM weights & $0.48$ \\ \hline
\multirow{2}{*}{Laplacian pseudoinverse (no learning, baseline)} 
    & \multirow{2}{*}{---} & Weighted Gaussian graph ($\eta=8$) & \multirow{2}{*}{---} & $0.01$ \\ \cline{3-3} \cline{5-5}
    &  & GNM ($\epsilon=10\angstrom$) &  & $0.01$ \\ \hline
\end{tabular}}
\label{table:S1}
\end{table}

\begin{figure}[H]
    \centering  
    \includegraphics[clip, trim=0cm 18cm 9cm 0cm, width=\linewidth]{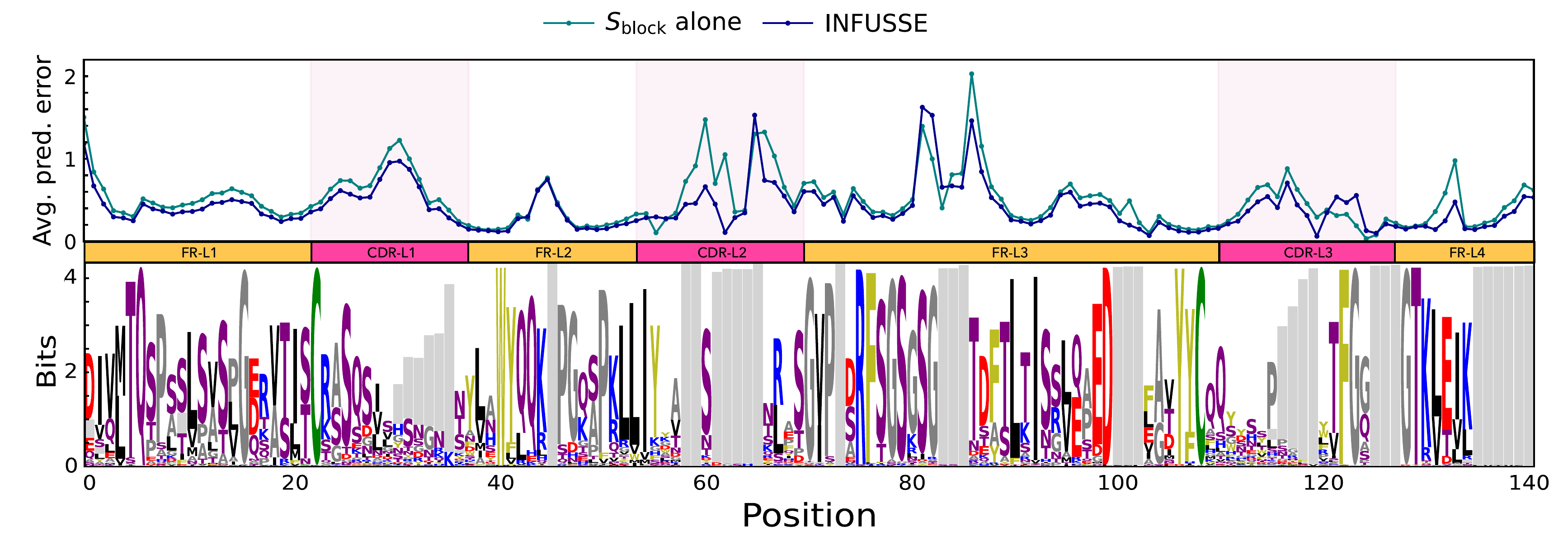}
  \caption{\textbf{INFUSSE’s predictions for the light chain variable region.} Same plot as Figure 2A, but for the light chain variable region. 
  }
    \label{fig:S1}
\end{figure}

\begin{figure}[H]
    \centering  \includegraphics[width=0.7\linewidth]{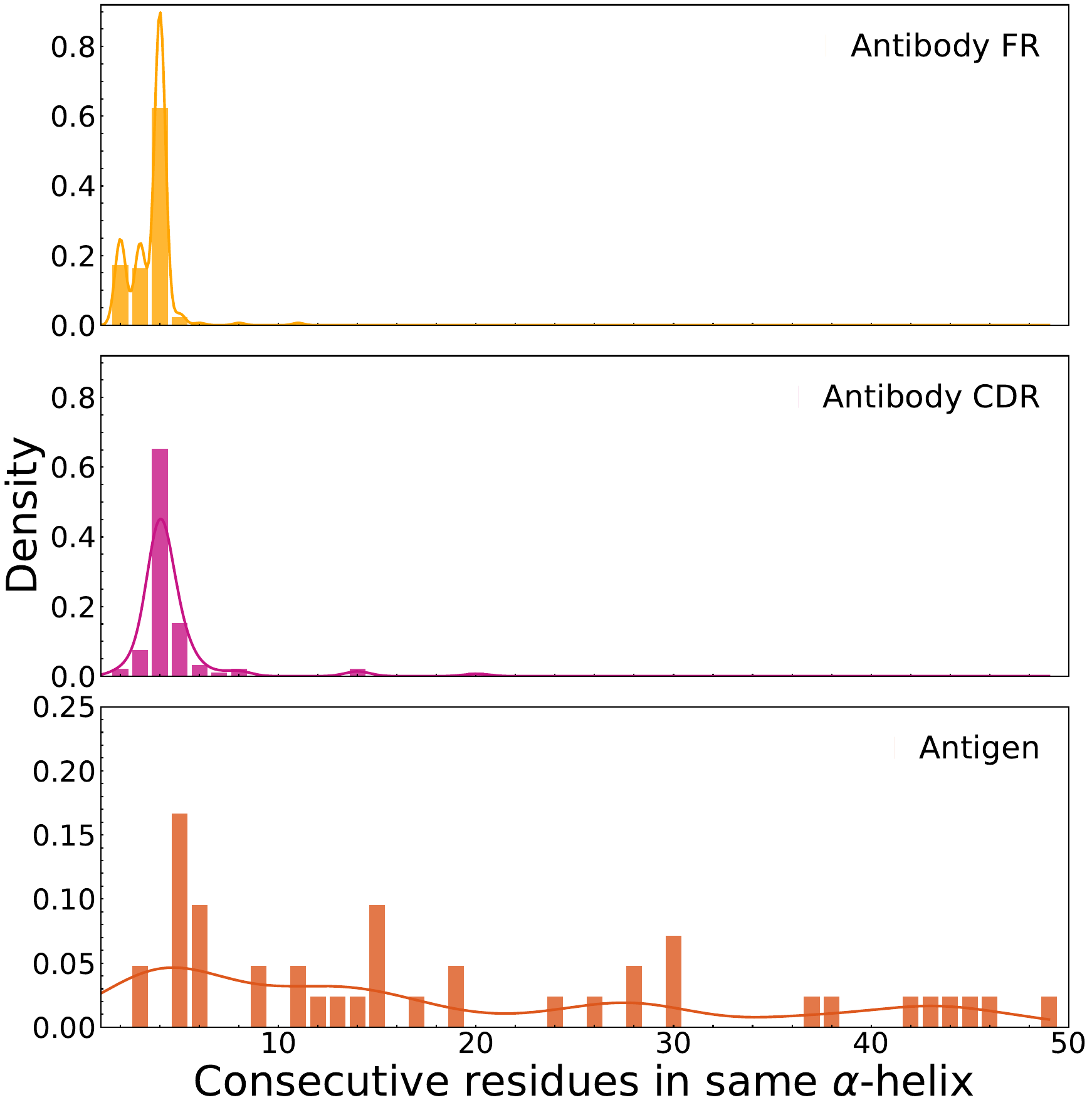}
  \caption{\textbf{Density of consecutive residues classed as being part of an $\alpha$-helix for the test set.} Histograms depict the distribution of consecutive $\alpha$-helix residues for antibody framework (FR), antibody complementarity-determining region (CDR) and antigen. Overlaid are Gaussian Kernel Density Estimator (KDE) curves, providing an estimation of the probability density function.
  }
    \label{fig:S2}
\end{figure}

\begin{figure}[H]
    \centering  \includegraphics[width=\linewidth]{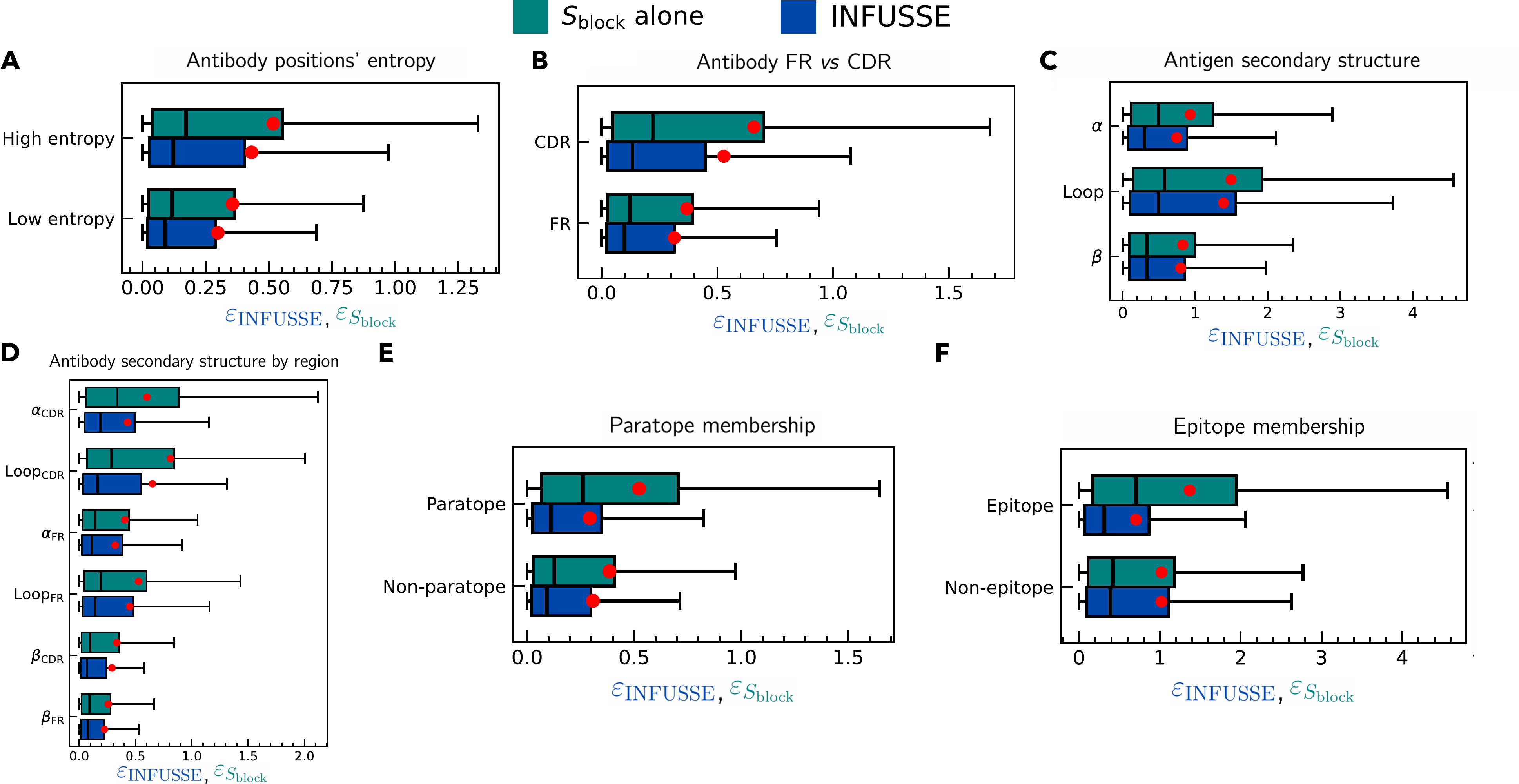}
  \caption{\textbf{Prediction errors of INFUSSE, $\varepsilon_{\text{INFUSSE}, j}^{(q)}$, and $S_{\mathrm{block}}$ alone, $\varepsilon_{S_\mathrm{block}, j}^{(q)}$, for each position $j$ and test set sample $q$.} The groups are based on: (\textbf{A}) antibody variable-region entropy (similar to \figref[C]{2}); (\textbf{B}) antibody framework (FR) versus complementarity-determining region (CDR) (similar to \figref[A]{3}); (\textbf{C}) antigen secondary structure (similar to \figref[B]{3}); (\textbf{D}) antibody region and secondary structure (similar to \figref[C]{3}); (\textbf{E}) paratope membership (similar to \figref[B]{4}); (\textbf{F}) epitope membership (similar to \figref[C]{4}). Red dots indicate mean values and black vertical lines denote medians.}
    \label{fig:S3}
\end{figure}

\begin{table}[H]
  \caption{
\textbf{Summary of statistical tests on the $\Delta_{\mathrm{graph}}$ distribution across different scenarios.} Upper-triangle cells show the difference in mean $\Delta\mu$ between the two groups and the p-value, lower-triangle cells the corresponding IQR differences $\Delta \mathrm{IQR}$ and the p-value; diagonal elements give the mean over the corresponding group (denoted by $\mu$) and its p-value. The groups compared are based on: (\textbf{A}) antibody variable-region entropy (data shown in~\figref[C]{2}); (\textbf{B}) antibody framework (FR) versus complementarity-determining region (CDR) (data shown in~\figref[A]{3}); 
(\textbf{C}) antigen secondary structure (data shown in~\figref[B]{3}); (\textbf{D}) antibody region and secondary structure (data shown in~\figref[C]{3}); 
(\textbf{E}) paratope membership (data shown in~\figref[B]{4}); (\textbf{F}) epitope membership (data shown in~\figref[C]{4}).  Statistical tests are carried out by the bootstrap as described in Methods, Section \ref{sec:statistical_tests}.}

\includegraphics[width=\linewidth,clip, trim=0cm 4cm 13.8cm 0cm]{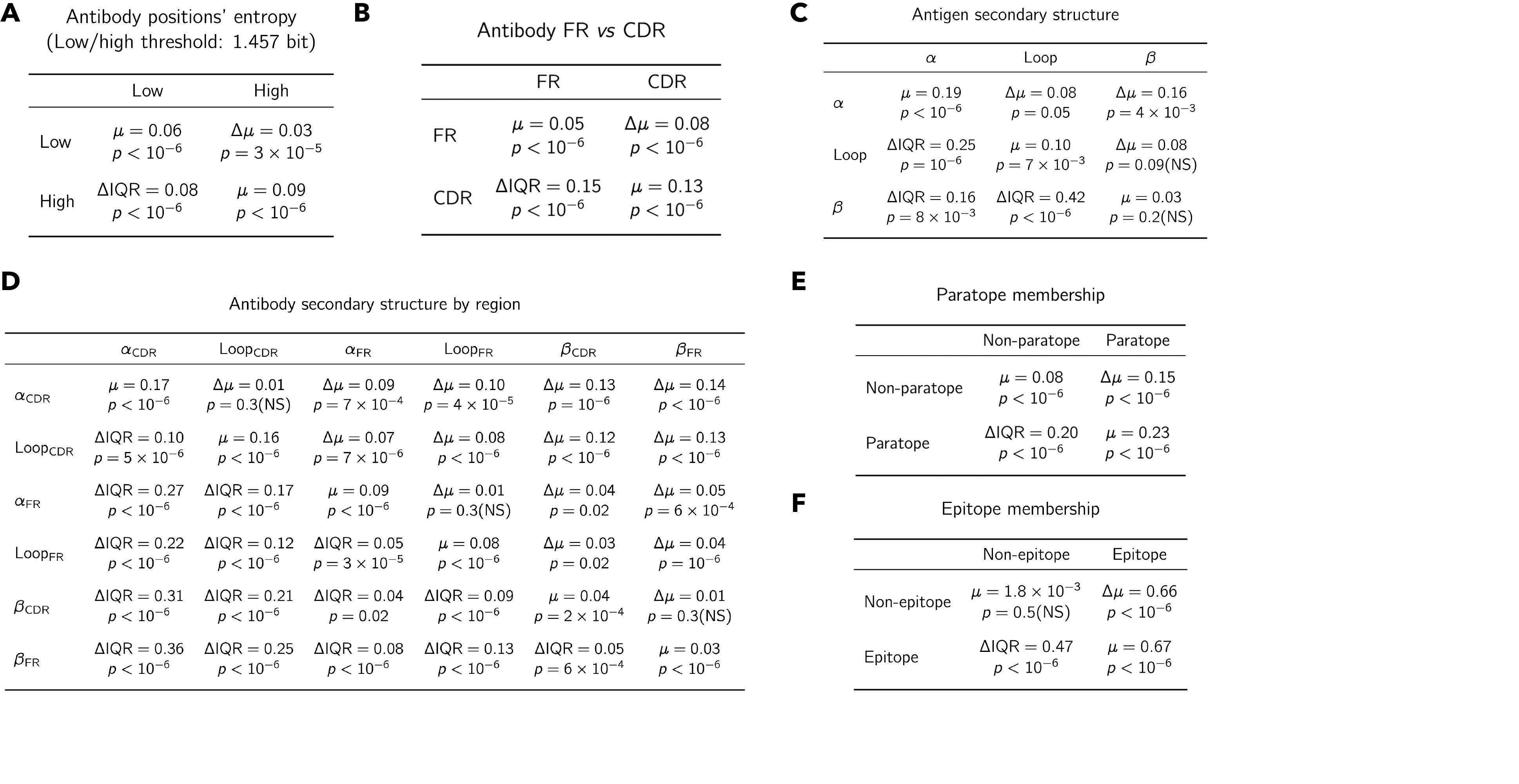}
  
    \label{table:S2}
\end{table}

\begin{table}[H]
  \caption{
\textbf{Summary of statistical tests on the $\varepsilon_{S_{\mathrm{block}}}$ distribution across different scenarios.} Panels and cells are the same as those in Table~\ref{table:S2}, while data is presented in Figure~\ref{fig:S3}. Statistical tests are carried out by the bootstrap as described in Methods, Section \ref{sec:statistical_tests}.}

\includegraphics[width=\linewidth,clip, trim=0cm 3.7cm 13.8cm 0cm]{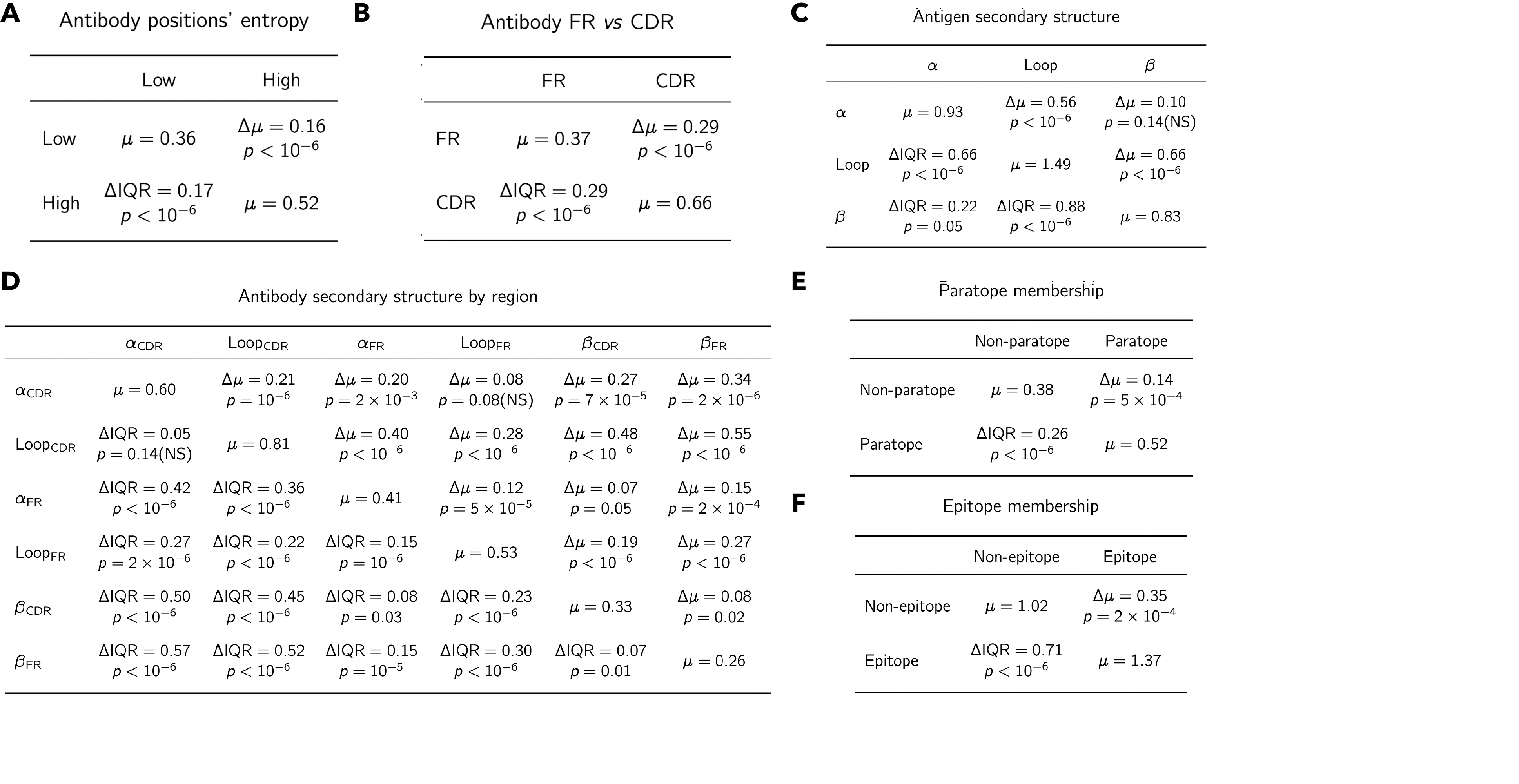}
  
    \label{table:S3}
\end{table}

\begin{table}[H]
  \caption{
\textbf{Summary of statistical tests on the $\varepsilon_{\mathrm{INFUSSE}}$ distribution across different scenarios.} Panels and cells are the same as those in Table~\ref{table:S2}, while data is presented in Figure~\ref{fig:S3}.  Statistical tests are carried out by the bootstrap as described in Methods, Section \ref{sec:statistical_tests}.}

\includegraphics[width=\linewidth,clip, trim=0cm 4cm 13.8cm 0cm]{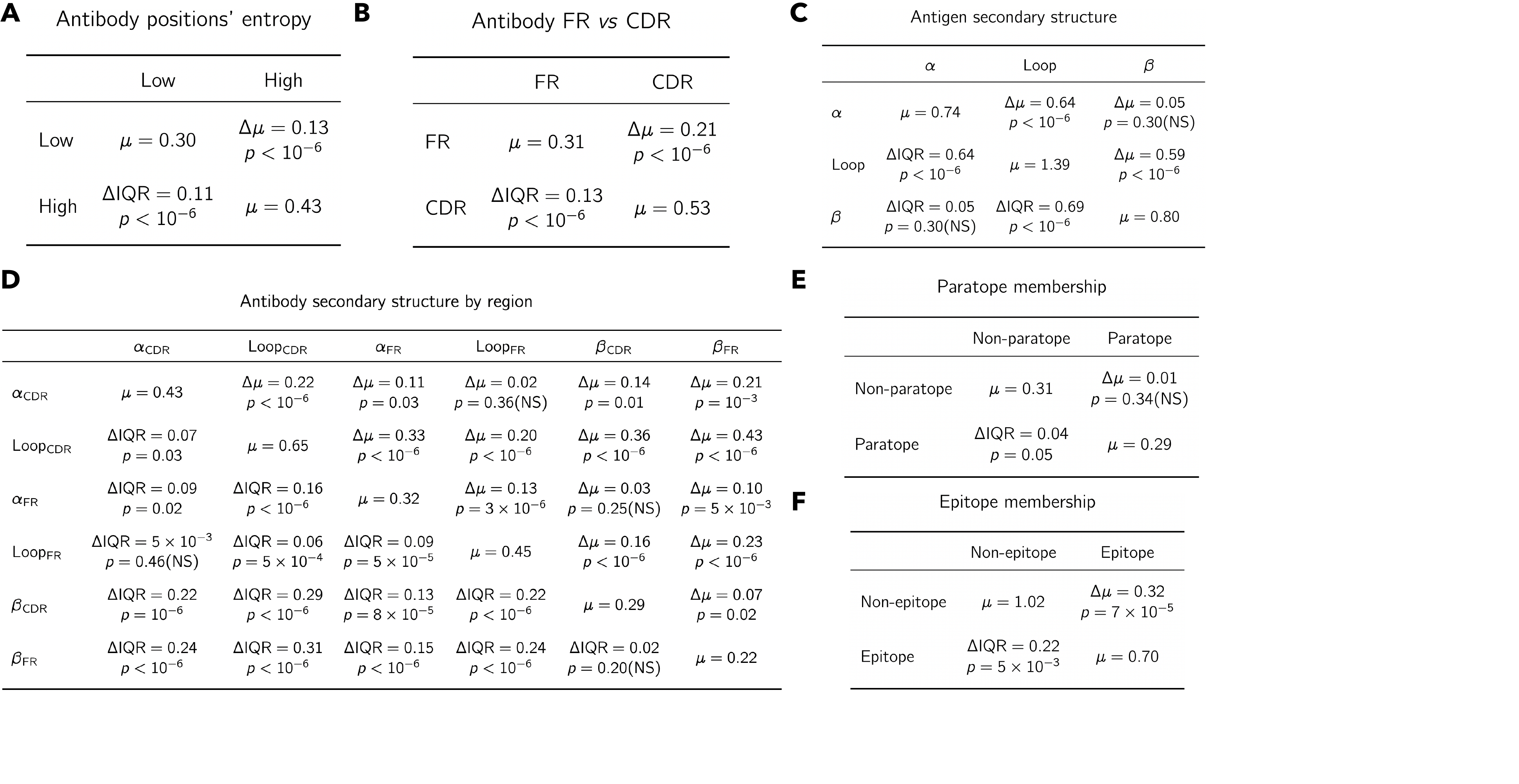}
  
    \label{table:S4}
\end{table}

\end{document}